\documentclass{article}%
\usepackage{amsmath}
\usepackage{graphicx}
\usepackage{amsfonts}
\usepackage{amssymb}
\usepackage{geometry}%
\providecommand{\U}[1]{\protect\rule{.1in}{.1in}}

\DeclareFontFamily{OT1}{pzc}{}
\DeclareFontShape{OT1}{pzc}{m}{it}{<-> s * [1.10] pzcmi7t}{}
\DeclareMathAlphabet{\mathpzc}{OT1}{pzc}{m}{it}
\providecommand{\U}[1]{\protect\rule{.1in}{.1in}}
\providecommand{\U}[1]{\protect\rule{.1in}{.1in}}
\providecommand{\U}[1]{\protect\rule{.1in}{.1in}}
\providecommand{\U}[1]{\protect\rule{.1in}{.1in}}
\providecommand{\U}[1]{\protect\rule{.1in}{.1in}}

\newcommand{\bee}{\begin{equation}}
\newcommand{\eend}{\end{equation}}

\newcommand{\bea}{\begin{eqnarray}}
\newcommand{\eea}{\end{eqnarray}}

\begin{document}

\title{\textbf{Nonlinear corrections in basic problems of electro- and
magneto-statics in the vacuum.}}
\author{Caio V. Costa$^{1}$\thanks{Electronic address: caiocostalopes@usp.br}, Dmitry
M. Gitman$^{1}$\thanks{Electronic address: gitman@dfn.if.usp.br} and Anatoly
E. Shabad$^{2}$\thanks{Electronic address: ashabad@Ipi.ru}\\$^{1}$\textsl{Instituto de F\'{\i}sica, Universidade de S\~{a}o Paulo, }\\\textsl{Caixa Postal 66318, CEP 05508-090, S\~{a}o Paulo, S. P., Brazil} \\$^{2}$\textsl{P. N. Lebedev Physics Institute, Leninsky Prospekt 53, Moscow
117924, Russia} }
\maketitle

\begin{abstract}
We find third-power nonlinear corrections to the Coulomb and other static
electric fields, as well as to the electric and magnetic dipole fields, as we
work within QED with no background field. The nonlinear response function we
base our consideration on is the fourth-rank polarization tensor, calculated
within the local (infrared) approximation of the effective action. Therefore,
the results are applicable to weakly varying fields. It is established that
the nonlinear correction to magnetic moment of some baryons just matches, in
the order of magnitude, the existing gap between its experimental and
theoretical values.

\end{abstract}

\bigskip

\bigskip

\section{Introduction}

\bigskip

Quantum electrodynamics (QED) is a nonlinear theory that includes effective
interaction between electromagnetic fields realized by creation of virtual
pairs of charged particles, electrons and positrons, that interact with the
electromagnetic fields before annihilating. The nonlinearity becomes usually
essential when the electromagnetic fields involved in the problem reach and
exceed the characteristic "Schwinger's" value of the order of $m^{2}/e,$ where
$m$ and $e$ are electron mass and charge\footnote{The only known exception is
provided by the resonance \cite{lettNuovCim} in the vacuum polarization
responsible for the capture of the photon \cite{nature} in pulsar
magnetospheres, where the photon forms a mixed state with the mutually bound
e$^{+}$e$^{-}$ pair \cite{Ass, Harding}. This effect is essential at the
magnetic field values of already $\approx0.1B_{\text{Sch}}.$} .\ In the Gauss
system of units this value is $B_{\text{Sch}}=4.4\times10^{13}$ G for magnetic
and $E_{\text{Sch}}=1.3\times10^{16}$ V/cm for electric field, while in the
Heaviside-Lorenz (HL) system, mostly used in the present paper, these values
are $1.2\times10^{13}$ G and $0.37\times10^{16}$ V/cm.

Apart from the customary reference to sufficiently large magnetic fields that
existed in the early Universe \cite{Vachaspati} or are existing \cite{Harding,
pulsars} in pulsars and magnetars (to be discussed in Section 5 below), it is
lately noted that the electric fields above Schwinger's value are expected
\cite{Co and Usov} to occur in quark stars, and that very strong magnetic
field is also formed for a short time when accelerated charged particles
(heavy ions) collide provided that the impact parameter is nonzero
\cite{Kharzeev}. Once the magnetic field arising in the collision is of the
order of hadron mass-squared and is thus apt of interfering with strong
interactions, this circumstance already avoked a vivid activity in QCD
\cite{Simonov}, lattice calculations \cite{Polikarpov} included.

Presently, we have also to note that large electric and magnetic fields may be
associated with baryons, due to their charges and/or to their electric and
magnetic multipole moments. Generally, large fields occur in the vicinity of
charged elementary particles or paricles carrying a magnetic or electric
moment, or both. Where these are protons, neutrons and, generally, atomic
nuclei, the stemming nonlinearity may affect the atomic spectra. (Within a
nonlinear electrodynamics theory, other than QED, namely, the noncommutative
$U_{\star}(1)$-gauge theory, this issue was touched in \cite{Stern}.)
Consider, for instance, the neutron, whose magnetic moment $\mathcal{M}$ is of
the order of $2$ nuclear magnetons $\mu_{N}=5.05\cdot10^{-24}%
\operatorname{g}%
^{1/2}$ $%
\operatorname{cm}%
^{5/2}%
\operatorname{s}%
^{-1}.$ The magnetic dipole field at the distance $r$ is, at its maximum,
$\sim2\mathcal{M}/r^{3}.$ It reaches Schwinger's value $4.4\cdot10^{\emph{13}}%
\operatorname{g}%
^{1/2}$ $%
\operatorname{cm}%
^{-1/2}%
\operatorname{s}%
^{-1}$ at the distance an order of magnitude larger than the neutron magnetic
size of $\sim1$ $%
\operatorname{fm}%
$, therefore the nonlinear correction to the nucleon magnetic moment
interacion with the orbital electron comes into play at the distance of about
five thousandth of the Bohr radius, which may be significant in the atomic scale.

\ \ There is a vast literature on nonlinear effects in QED, where the
electromagnetic field, which is necessary to be taken as large is the so
called external field \ that forms the vacuum background, see, \textit{e.g}.,
the monographs \cite{trudy, Gies, Kuznetsov}. In this approach, a great
advantage of going beyond the perturbation expansions in powers of the
external field is achieved thanks to the use of special fields that admit
exact solutions for the Dirac electron Green function, like fields constant in
space and time and the plane wave (laser) field. In the present, as well as in
the previous \cite{GitShab}, paper we are interested also in another class of
fields, namely those that are produced by spherically and cylindrically
symmetric static sources, including the fields of electric and magnetic
dipole. As far as exact Dirac solutions for such fields are either unknown or
overcomplicated to exploit, we treat them relying on the power expansion and
take into account the lowest nontrivial power. Their strength should be enough
to make the contribution of nonlinearity significant, but still sufficiently
small to keep below the value, where the power expansion remains meaningful.
On this basis we considered previously the magnetic field produced by a static
charge placed in a strong constant and homogeneous magnetic field
\cite{GitShab}. The latter was treated as the nonperturbative background, it
might exceed the Schwinger value, whereas the contribution of the static
charge was retained up to the second power. The very existence of this
contribution (the magneto-electric effect in QED) is provided by the
three-photon diagram (third-rank polarization tensor) off the photon mass
shell, nonzero against the external background. Now we switch off the
background field and confine ourselves to various static sources in the blank
vacuum. This time, the third-rank polarization tensor vanishes due to the
C-invariance, and the first nonlinear self-coupling of the sources originates
from the fourth-order polarization tensor associated with the four-photon
diagram. This diagram, when reduced to the mass shell of four or two photons,
describes the photon-by-photon scattering and scattering of a photon off a
Coulomb center, respectively. This exhausts all scattering processes with the
four-prong diagram \cite{Akhiezer}. However, when all the four photon legs,
are taken off-shell, i.e., beyond the photon dispersion curves, this diagram
can be used for calculation of self-couplings of various sources of
electromagnetic fields. We present a framework for considering slowly varying
in time and space sources on the basis of effective action formalism, where
the fourth-rank polarization tensor is calculated by four variational
field-derivatives of the effective action taken as a local functional of the
field invariants. This corresponds, in momentum space, to the infrared limit
of small 4-momenta. This framework supplies us with the simplest possible
approximation, wherein the self-coupling becomes first pronounced. For a
point-like electric charge the nonlinear correction to the Coulomb field
presented in \cite{BerLifPit} (commented on also in \cite{GitShab}) is
reproduced. Extended electric and magnetic sources of various symmetries are
studied, and nonlinear corrections to the fields produced by them are found,
the most attention paid to the electric and magnetic dipole fields. These
fields reproduce themselves under the mapping due to the nonlinearity, while
the resulting magnetic and electric moments undergo nonlinear corrections,
becoming subject of cubic equations.

Our numerical estimates show that the fields of giant magnetic dipoles, the
pulsars \cite{Beskin}, do produce essential nonlinear corrections, before they
reach the values of the order of $10^{15}%
\operatorname{G}%
,$ when we are already taken out of the scope of validity of our power
expansion. The nonlinear corrections can hardly be distinguished by a
far-remote observer, since these lead only to renormalization of the magnetic
moment. However, inside and in the close vicinity of magnetars the nonlinear
corrections to any theoretically derived model values of the magnetic field
are huge, and must be taken into account when considering various physical
processes in these regions.

As far as baryons are concerned we find that the magnetic moment of the
neutron is too large to be treated within the present power-expansion
approach. Nevertheless, the nonlinear corrections to smaller magnetic moments
of some other baryons do fit this approach and prove to be of the same order
of magnitude as the existing indeterminacy in the theoretical results for
their magnetic moment values calculated within the theory of strong
interactions. This means that the nonlinearity of QED will have to be taken
into account already at the very next step of perfection of the theoretical
predictions for the magnetic moments of these neutral baryons.

Before proceeding, we find it interesting to note in passing that the
magnetization of the order of extreme magnetars may be associated with the
magnetic moment of the neutron. If one imagines a neutron star as composed of
neutrons in a ferromagnetic state, i.e. with all magnetic moments of the
constituent neutrons parallel to each other, and packed as densely as their
electromagnetic size permits, the magnetization of that star would be the same
as the magnetization of an individual neutron, defined as its magnetic moment
per unit volume. This is about (see Subsection 5.2 below) $3\times10^{16}$ $%
\operatorname{G}%
.$ This numerical observation may support the idea of essential nucleon
contribution (developed, \textit{e.g.,} in \cite{hadron}) into the magnetic
field of neutron stars (besides the notion of electron currents) and of the
magnetar value.

\ The paper is organized as follows. In Section 2 for any U(1)-
gauge-invariant P-even theory we present the nonlinear Maxwell equations for
field potentials to the fourth-power accuracy in gauge fields . The second-,
third- and fourth-rank polarization tensors are calculated in Appendix1 by
differetiations of the effective action over the external constant field to be
set equal to zero afterwards. When there is no background, only the
third-power nonlinear vacuum response, contained in the fourth-order
field-derivative of the effective action, survives the infrared limit. The
quadratic nonlinear response is absent, as noted above. As for the linear
response, it is trivial in the zero-momentum limit: the second-rank
polarization tensor vanishes due to the correspondence principle. We define a
nonlinearly induced current as the source of the nonlinear correction to the fields.

In Section 3 we derive a representation for the nonlinear field strengths in
special cases where either only magnetic or electric field is present. These
representations contain projection operators. In Section 4, Subsection 4.1,
the highly symmetric configurations are considered, for which these projection
operators are identities, and the fields are mapped to their nonlinear
counterparts in a local way. In this context nonlinear corrections to
spherically symmetric extended charges, charged thread, charged plane of
finite thickness and a wire with current are found. More complicated
cylindrically symmetric dipole fields produced by a current and by a charge
distributed in a special way over the surface of a sphere are studied in
Subsection 4.2. The details of calculations intended to realize the projection
operators are given in Appendix 2 (magnetic dipole) and Appendix 3 (electric
dipole). The resulting equations for self-coupling of dipole moments are
discussed. We found that there is no spontaneous creation of these quantities
thanks to special property of the second derivative of the effective action
that reflects the causality and unitarity properties of the effective action.
At the last step the Euler-Heisenberg (local) effective action is used to
specify the results dynamically to QED.

In Section 5 we make numerical estimates of the results as applied to magnetic
stars and baryons. The contents is summarized in concluding Section 6.

We adhere to the rationalized Heaviside system of units with $\hbar
=c=\varepsilon_{0}=1$ throughout, unless the opposite is indicated.

\section{Nonlinear electromagnetic field equations}

We describe the effective action
\begin{equation}
\Gamma=\int\mathfrak{L}\left(  z\right)  \mathrm{d}^{4}z, \label{Gamma}%
\end{equation}
in QED as the Legendre transform of the generating functional of the Green
functions \cite{weinberg}. It is, in its turn, the generating functional of
the one-particle-irreducible vertices. Eq. (\ref{Gamma}) is a nonlocal
functional, $\mathfrak{L}$ being the effective Lagrangian, which depends on
the relativistic field invariants $\mathfrak{F}\left(  x\right)  =\frac{1}%
{4}F_{\mu\nu}\left(  x\right)  F^{\mu\nu}\left(  x\right)  $, $\mathfrak{G}%
\left(  x\right)  =\frac{1}{4}F_{\mu\nu}\left(  x\right)  \tilde{F}^{\mu\nu
}\left(  x\right)  $ and, generally, on their space-time derivatives of any
order. Here the field strength is $F_{\mu\nu}\left(  x\right)  =\partial_{\mu
}A_{\nu}\left(  x\right)  -\partial_{\nu}A_{\mu}\left(  x\right)  $,
with$\ A_{\mu}\left(  x\right)  $ being the field potentials, and $\tilde
{F}_{\mu\nu}\left(  x\right)  =\frac{1}{2}\epsilon_{\mu\nu\alpha\beta
}F^{\alpha\beta}\left(  x\right)  $ is the dual tensor, where $\epsilon
_{\mu\nu\alpha\beta}$ is the fully antisymmetric Levi-Civita tensor, such that
$\epsilon_{0123}=1.$ The Greek indices span the 4-dimensional Minkowski space
taking the values $0,1,2,3$. The metric tensor is $\left(  \eta_{\mu\tau
}\right)  =\mathrm{diag}\left(  1,-1,-1,-1\right)  $, and latin indices take
the values $1,2,3$.

Beyond QED the action (\ref{Gamma}) should be thought of a primarily given
classical action of a nonlinear theory,\textit{ e.g.}, the Born-Infeld action
or its nonlocal generalization. As a matter of fact the contents of this
article, down to the very last step, where we refer to the Euler-Heisenberg
Lagrangian, is independent of special dynamcs: it covers QED as well as any
other nonlinear electrodynamcs.

Once we want to find and solve the Euler-Lagrange equations, which the
potentials $A^{\mu}\left(  x\right)  $ must satisfy, given some current
$J_{\beta}\left(  x\right)  $, then we need to apply the variational principle
to the total action $S_{tot}$, defined as%
\begin{align}
S_{\mathrm{tot}}  &  =S+S_{\mathrm{int}},\text{ }S=S_{\mathrm{free}}%
+\Gamma\nonumber\\
S_{\mathrm{free}}  &  =-\int\mathfrak{F}\left(  z\right)  \mathrm{d}%
^{4}z,\text{ }S_{\mathrm{int}}=-\int J_{\beta}\left(  z\right)  A^{\beta
}\left(  z\right)  \mathrm{d}^{4}z. \label{2}%
\end{align}
The variational principle imposes minimum action on $S_{\mathrm{tot}}$, i.e.,
$\frac{\delta S_{\mathrm{tot}}}{\delta A^{\beta}\left(  x\right)  }=0$,
therefore we get the Euler-Lagrange equations%
\begin{equation}
J_{\mu}\left(  z\right)  =\frac{\delta S_{\mathrm{free}}}{\delta A^{\mu
}\left(  x\right)  }+\frac{\delta\Gamma}{\delta A^{\mu}\left(  x\right)  }.
\label{Euler-Lagrange equations}%
\end{equation}
They provide exact nonlinear equations for the c-valued electromagnetic field
that include all quantum corrections inherent in QED. The linear part of these
equations makes the standard Maxwell equations, valid in the small-field
limit, in an equivalent medium formed by the polarized vacuum with or without
external field background.

Note that when writin Eqs. (\ref{2}), (\ref{Euler-Lagrange equations}) we keep
to HL system, since the energy of of the free field and the Euler-Lagrange
(Maxwell) equations do not contain the factor $1/4\pi.$

\subsection{Expansion in powers of the field above the background}

Subdivide the current in two parts: $J_{\beta}\left(  x\right)  =j_{\beta
}\left(  x\right)  +\mathcal{J}_{\beta}\left(  x\right)  $, where
$\mathcal{J}_{\beta}\left(  x\right)  $ is the current supporting an external
field $\mathcal{A}^{\beta}\left(  x\right)  $ via the exact nonlinear Maxwell
equation $\mathcal{J}_{\beta}\left(  x\right)  $ $=\left.  \frac{\delta
S}{\delta A^{\beta}\left(  x\right)  }\right\vert _{A=\mathcal{A}}=\left[
\square\eta_{\beta\tau}-\partial^{\mathcal{\beta}}\partial^{\tau}\right]
\mathcal{A}^{\tau}\left(  x\right)  +\left.  \frac{\delta\Gamma}{\delta
A^{\beta}\left(  x\right)  }\right\vert _{A=\mathcal{A}}$. We shall seek
solutions in the form $A^{\beta}\left(  x\right)  =\mathcal{A}^{\beta}\left(
x\right)  +a^{\beta}\left(  x\right)  $. The current $j_{\beta}\left(
x\right)  $ is supposed to be small, and then we can expand $\frac
{\delta\Gamma}{\delta A^{\beta}\left(  x\right)  }$ in powers of the small
field $a^{\beta}\left(  x\right)  $ above the external field background.
Hence, (\ref{Euler-Lagrange equations}) becomes
\begin{align}
&  j_{\mu}\left(  x\right)  =\left[  \square\eta_{\mu\tau}-\partial_{\mu
}\partial_{\tau}\right]  a^{\tau}\left(  x\right)  +\int\mathrm{d}^{4}y\text{
}\Pi_{\mu\tau}\left(  x,y\right)  a^{\tau}\left(  y\right)
\label{General current}\\
&  +\frac{1}{2}\int\mathrm{d}^{4}y\mathrm{d}^{4}u\text{ }\Pi_{\mu\tau\sigma
}\left(  x,y,u\right)  a^{\tau}\left(  y\right)  a^{\sigma}\left(  u\right)
\nonumber\\
&  +\frac{1}{6}\int\mathrm{d}^{4}y\mathrm{d}^{4}u\mathrm{d}^{4}v\text{ }%
\Pi_{\mu\tau\sigma\rho}\left(  x,y,u,v\right)  a^{\tau}\left(  y\right)
a^{\sigma}\left(  u\right)  a^{\rho}\left(  v\right)  ,\nonumber
\end{align}
where $\square=\partial_{0}^{2}-\boldsymbol{\nabla}^{2}$, and we have
restricted ourselves to the third-power terms in the expansion. The expansion
parameter can be revealed no sooner than the matrix coefficients of the
expansion (2-, 3- and 4-rank polarizatioin tensors)
\begin{equation}
\Pi_{\mu\tau}\left(  x,x^{\prime}\right)  =\left.  \frac{\delta^{2}\Gamma
}{\delta A^{\mu}\left(  x\right)  \delta A^{\tau}\left(  x^{\prime}\right)
}\right\vert _{A=\mathcal{A}}, \label{Pi}%
\end{equation}%
\begin{equation}
\Pi_{\mu\tau\sigma}\left(  x,x^{\prime},x^{\prime\prime}\right)  =\left.
\frac{\delta^{3}\Gamma}{\delta A^{\mu}\left(  x\right)  \delta A^{\tau}\left(
x^{\prime}\right)  \delta A^{\sigma}\left(  x^{\prime\prime}\right)
}\right\vert _{A=\mathcal{A}}, \label{Pi3}%
\end{equation}%
\begin{equation}
\Pi_{\mu\tau\sigma\rho}\left(  x,x^{\prime},x^{\prime\prime},x^{\prime
\prime\prime}\right)  =\left.  \frac{\delta^{4}\Gamma}{\delta A^{\mu}\left(
x\right)  \delta A^{\tau}\left(  x^{\prime}\right)  \delta A^{\sigma}\left(
x^{\prime\prime}\right)  \delta A^{\rho}\left(  x^{\prime\prime\prime}\right)
}\right\vert _{A=\mathcal{A}} \label{Pi4}%
\end{equation}
are calculated within one or another dynamical scheme. The polarization
tensors of every rank \ $\Pi_{\mu\tau...\sigma}(x,x^{\prime},...x^{\prime
\prime})$\ satisfy the continuity relations with respect to every argument and
every index as a consequence of the gauge invariance.

In this work, we shall mainly deal with external fields $\mathcal{F}%
_{\alpha\beta}\mathcal{=\partial}^{\alpha}\mathcal{A}_{\beta}^{\text{ext}%
}\mathcal{-\partial}^{\beta}\mathcal{A}_{\alpha}^{\text{ext}}$ equal to zero
and call this background the blank vacuum. (The only exceptions are
calculations in Appendix 1, which include the field $\mathcal{F}_{\alpha\beta
}$\ constant in space and time within the framework explained in
\cite{GitShab}). In this case (also for the constant background) all-rank
polarization tensors depend on their coordinate differences. Eq.
(\ref{General current}) is the field equation for small electromagnetic
perturbations $a^{\beta}(x)=A^{\beta}\left(  x\right)  -\mathcal{A}^{\beta
}(x)$ over the blank vacuum, caused by a small external current $j_{\rho}(x)$
and taken to the lowest-power nonlinearity.

With the definition of the photon propagator $D_{\mu\nu}(x,x^{\prime})$%
\begin{equation}
D_{\mu\nu}^{-1}(x-x^{\prime})=\left[  \eta_{\mu\nu}\square-\partial^{\mu
}\partial^{\nu}\right]  \delta^{(4)}(x^{\prime}-x)+\Pi_{\mu\nu}(x-x^{\prime})
\label{propagator}%
\end{equation}
the nonlinear field equations (\ref{General current}) take the form of (the
set of) integral equations%

\begin{equation}
a^{\lambda}(x)=\int d^{4}yD^{\lambda\rho}(x-y)j_{\rho}(y)+\int d^{4}%
yD^{\lambda\rho}(x-y)j_{\rho}^{\text{\textrm{nl}}}(y), \label{a}%
\end{equation}

\begin{align}
&  j_{\mu}^{\text{\textrm{nl}}}(x)=-\frac{1}{2}\int\mathrm{d}^{4}%
y\mathrm{d}^{4}u\text{ }\Pi_{\mu\tau\sigma}\left(  x,y,u\right)  a^{\tau
}\left(  y\right)  a^{\sigma}\left(  u\right) \label{nonlincur}\\
&  -\frac{1}{6}\int\mathrm{d}^{4}y\mathrm{d}^{4}u\mathrm{d}^{4}v\text{ }%
\Pi_{\mu\tau\sigma\rho}\left(  x,y,u,v\right)  a^{\tau}\left(  y\right)
a^{\sigma}\left(  u\right)  a^{\rho}\left(  v\right)  ,\nonumber
\end{align}
where we define $j_{\mu}^{\text{\textrm{nl}}}(x)$, and call it "nonlinearly
induced current"$.$

\subsection{Infrared approximation and its application in no-external-field
vacuum}

From now on we shall restrict ourselves only to slowly varying fields
$a^{\lambda}(x)$ and, correspondingly, to consideration of the sources
$j_{\rho}(y)$ that give rise to such fields via equations (\ref{a}),
(\ref{nonlincur}). To this end we may take the effective action $\Gamma$ in
the local limit, where the field derivatives are disregarded from this
functional. We also call this limit infrared, because the variational
derivatives for the $n$-rank polarization operators (\ref{Pi}), (\ref{Pi3}),
(\ref{Pi4}) become in the local limit their low-momentum asymptotes $\sim
k^{n},$ where the momentum $k_{\mu}$ is the variable, Fourier-conjugate to
$x_{\mu}.$The resulting expressions for (\ref{Pi}), (\ref{Pi3}), (\ref{Pi4})
with the background fields $\mathcal{F}_{\alpha\beta}$ $\mathcal{=\partial
}^{\alpha}\mathcal{A}_{\beta}^{\text{ext}}\mathcal{-\partial}^{\beta
}\mathcal{A}_{\alpha}^{\text{ext}}$ being arbitrary combinations of constant
and homogeneous electric and magnetic fields are calculated in the infrared
limit following the same method as in \cite{GitShab}, and listed in Appendix
1, Eqs. (\ref{secdir}), (\ref{thirdir}), (\ref{Fourth rank tensor}).
Henceforward, however, we set the background field $\mathcal{F}_{\alpha\beta}$
equal to zero everywhere, because we are interested only in the blank vacuum
as the background medium in the present work, leaving the $\mathcal{F}%
_{\alpha\beta}\neq0$ calculations as a billet for a future use. Then the
variational derivatives in (\ref{Pi}), (\ref{Pi3}) and (\ref{Pi4}) are
expressed in terms of derivatives of $\mathfrak{L}\left(  \mathfrak{F,G}%
\right)  $ with respect to the field invariants reduced to the null external
field $\mathcal{A=}$ $0$.

Define $\mathfrak{L}_{\mathfrak{F}}=\left.  \frac{\partial\mathfrak{L}%
}{\partial\mathfrak{F}}\right\vert _{\mathfrak{F}=0,\mathfrak{G}=0}$,
$\mathfrak{L}_{\mathfrak{G}}=\left.  \frac{\partial\mathfrak{L}}%
{\partial\mathfrak{G}}\right\vert _{\mathfrak{F}=0,\mathfrak{G}=0}%
,\mathfrak{L}_{\mathfrak{FF}}=\left.  \frac{\partial^{2}\mathfrak{L}}%
{\partial\mathfrak{F}^{2}}\right\vert _{\mathfrak{F}=0,\mathfrak{G}=0},$
$\mathfrak{L}_{\mathfrak{FG}}=\left.  \frac{\partial^{2}\mathfrak{L}}%
{\partial\mathfrak{F}\partial\mathfrak{G}}\right\vert _{\mathfrak{F}%
=0,\mathfrak{G}=0}$ and $\mathfrak{L}_{\mathfrak{GG}}=\left.  \frac
{\partial^{2}\mathfrak{L}}{\partial\mathfrak{G}^{2}}\right\vert _{\mathfrak{F}%
=0,\mathfrak{G}=0}$. The relations $\mathfrak{L}_{\mathfrak{F}}=\mathfrak{L}%
_{\mathfrak{G}}=0$ hold thanks to the correspondence principle, that reads
that in the infrared limit the standard linear Maxwell equations for the
no-background-field case should be exact, not subject to any corrections.
Another reason for $\mathfrak{L}_{\mathfrak{G}}$ (also for $\mathfrak{L}%
_{\mathfrak{FG}}$) to vanish is the P-invariance. Setting $\mathcal{F}%
_{\alpha\beta}=$ $\mathcal{\tilde{F}}_{\alpha\mu}$ = 0 in (\ref{secdir}),
(\ref{thirdir}), (\ref{Fourth rank tensor}) we make sure that the linear
response in the blank vacuum is trivial in the local (infrared)
limit\footnote{This consequence of the correspondence principle $\mathfrak{L}%
_{\mathfrak{F}}=0$ is seen in the structure of the renormalized polarization
operator in the momentum representation prescribed by the standard
renormalization procedure, see, \textit{e.g}., \cite{Akhiezer}, that respects
that principle, $\Pi_{\mu\tau}^{R}=\left(  \eta_{\mu\tau}k^{2}-k_{\mu}k_{\tau
}\right)  \left(  \Pi\left(  k^{2}\right)  -\Pi(0)\right)  :$ this goes to
zero as $\sim k^{4}$ in the infrared limit.} (and not beyond it, of course):%
\begin{equation}
\Pi_{\mu\tau}\left(  x,y\right)  =0 \label{Linear response}%
\end{equation}
while the quadratic response disappears:%

\begin{equation}
\Pi_{\mu\tau\sigma}\left(  x,y,u\right)  =0, \label{no quadratic response}%
\end{equation}
the latter property being as a matter of fact a consequence of C-invariance,
valid beyond the infrared approximation as well. As for the cubic response, it
is governed by the fourth-rank tensor (\ref{Pi4}), which in the current case
follows from (\ref{Fourth rank tensor}) to be%
\begin{align}
&  \Pi_{\mu\tau\sigma\rho}\left(  x,y,u,v\right)
=\label{Fourth rank polarizator tensor}\\
&  =\int\mathrm{d}^{4}z\left\{  \mathcal{P}_{\alpha\beta\gamma\lambda\mu
\tau\sigma\rho}\frac{\partial\delta^{4}\left(  x-z\right)  }{\partial
z_{\alpha}}\frac{\partial\delta^{4}\left(  y-z\right)  }{\partial z_{\beta}%
}\frac{\partial\delta^{4}\left(  u-z\right)  }{\partial z_{\gamma}}%
\frac{\partial\delta^{4}\left(  z-v\right)  }{\partial z_{\lambda}}\right\}
\text{,}\nonumber
\end{align}
where we define the constant tensor $\mathcal{P}_{\alpha\beta\gamma\lambda
\mu\tau\sigma\rho}$
\begin{align}
&  \mathcal{P}_{\alpha\beta\gamma\lambda\mu\tau\sigma\rho}=\mathfrak{L}%
_{\mathfrak{FF}}\left[  \left(  \eta_{\alpha\lambda}^{\text{ }}\eta_{\rho\mu
}-\eta_{\mu\lambda}^{\text{ }}\eta_{\alpha\rho}\right)  \left(  \eta
_{\tau\sigma}^{\text{ }}\eta_{\beta\gamma}-\eta_{\beta\sigma}^{\text{ }}%
\eta_{\tau\gamma}\right)  \right. \label{Constant tensor}\\
&  \left.  +\left(  \eta_{\beta\lambda}^{\text{ }}\eta_{\rho\tau}-\eta
_{\tau\lambda}^{\text{ }}\eta_{\beta\rho}^{\text{ }}\right)  \left(  \eta
_{\mu\sigma}^{\text{ }}\eta_{\alpha\gamma}-\eta_{\alpha\sigma}^{\text{ }}%
\eta_{\mu\gamma}\right)  +\left(  \eta_{\gamma\lambda}^{\text{ }}\eta
_{\rho\sigma}-\eta_{\sigma\lambda}^{\text{ }}\eta_{\gamma\rho}\right)  \left(
\eta_{\mu\tau}^{\text{ }}\eta_{\alpha\beta}-\eta_{\mu\beta}^{\text{ }}%
\eta_{\alpha\tau}^{\text{ }}\right)  \right] \\
&  +\mathfrak{L}_{\mathfrak{GG}}\left[  \epsilon_{\alpha\mu\beta\tau}%
\epsilon_{\lambda\rho\gamma\sigma}+\epsilon_{\lambda\rho\alpha\mu}%
\epsilon_{\beta\tau\gamma\sigma}+\epsilon_{\lambda\rho\beta\tau}%
\epsilon_{\alpha\mu\gamma\sigma}\right]  .\nonumber
\end{align}

When deriving this expression we have restricted ourselves to the P-even
theories, to which class QED belongs, by imposing the extra condition
$\mathfrak{L}_{\mathfrak{FG}}=0$. Integrating
(\ref{Fourth rank polarizator tensor}) by parts we obtain%
\begin{equation}
\Pi_{\mu\tau\sigma\rho}\left(  x,y,u,v\right)  =-\mathcal{P}_{\alpha
\beta\gamma\lambda\mu\tau\sigma\rho}\text{ }\frac{\partial}{\partial
x_{\alpha}}\left\{  \left[  \frac{\partial\delta^{4}\left(  y-x\right)
}{\partial x_{\beta}}\right]  \left[  \frac{\partial\delta^{4}\left(
u-x\right)  }{\partial x_{\gamma}}\right]  \left[  \frac{\partial\delta
^{4}\left(  x-v\right)  }{\partial x_{\lambda}}\right]  \right\}  .
\label{4 rank}%
\end{equation}

With the account of (\ref{Linear response}) and (\ref{no quadratic response}),
the Maxwell equations (\ref{General current}) including the cubic nonlinearity
reduce to%

\begin{equation}
j_{\mu}\left(  x\right)  =\left[  \square\eta_{\mu\tau}-\partial_{\mu}%
\partial_{\tau}\right]  a^{\tau}\left(  x\right)  -j_{\mu}^{\mathrm{nl}%
}\left(  x\right)  , \label{field equation}%
\end{equation}
where the nonlinearly induced current (\ref{nonlincur}) is reduced to%
\begin{equation}
j_{\mu}^{\mathrm{nl}}\left(  x\right)  =-\frac{1}{6}\int\mathrm{d}%
^{4}y\mathrm{d}^{4}u\mathrm{d}^{4}v\text{ }\Pi_{\mu\tau\sigma\rho}\left(
x,y,u,v\right)  a^{\tau}\left(  y\right)  a^{\sigma}\left(  u\right)  a^{\rho
}\left(  v\right)  . \label{Non linear induced current}%
\end{equation}
Divide the field into the sum of "linear" and "nonlinear" parts%
\begin{equation}
a^{\tau}\left(  x\right)  =a_{\text{lin}}^{\tau}\left(  x\right)
+a_{\text{nl}}^{\tau}\left(  x\right)  \label{linear+nonlinear}%
\end{equation}
so that Eq.(\ref{field equation}) becomes%
\begin{align}
j_{\mu}\left(  x\right)   &  =\left[  \square\eta_{\mu\tau}-\partial_{\mu
}\partial_{\tau}\right]  a_{\text{lin}}^{\tau}\left(  x\right)
,\label{linfield}\\
j_{\mu}^{\text{nl}}\left(  x\right)   &  =\left[  \square\eta_{\mu\tau
}-\partial_{\mu}\partial_{\tau}\right]  a_{\text{nl}}^{\tau}\left(  x\right)
\label{nonlinfield}%
\end{align}
In the forthcoming sections, to solve the cubic nonlinear Maxwell equations,
the set (\ref{field equation}), (\ref{Non linear induced current}), we will be
treating the nonlinearity by iterations. For a time being, however, we
continue to keep it as it is, i.e. to be expressing the nonlinearly induced
current $j_{\mu}^{\text{nl}}\left(  x\right)  $ in terms of the exact fields
$a^{\tau}\left(  x\right)  $ as given by (\ref{Non linear induced current}).

With the use of (\ref{4 rank}) and of integration by parts we find%
\begin{align*}
&  j_{\mu}^{\mathrm{nl}}\left(  x\right)  =\frac{\mathcal{P}_{\alpha
\beta\gamma\lambda\mu\tau\sigma\rho}}{6}\frac{\partial}{\partial x_{\alpha}%
}\left[  \int\text{ }\frac{\partial\delta^{4}\left(  y-x\right)  }{\partial
x_{\beta}}a^{\tau}\left(  y\right)  \mathrm{d}^{4}y\int\frac{\partial
\delta^{4}\left(  u-x\right)  }{\partial x_{\gamma}}a^{\sigma}\left(
u\right)  \mathrm{d}^{4}u\right. \\
\times &  \left.  \int\frac{\partial\delta^{4}\left(  x-v\right)  }{\partial
x_{\lambda}}a^{\rho}\left(  v\right)  \mathrm{d}^{4}v\right]  =\frac
{\mathcal{P}_{\alpha\beta\gamma\lambda\mu\tau\sigma\rho}}{6}\frac{\partial
}{\partial x_{\alpha}}\left[  \frac{\partial a^{\tau}\left(  x\right)
}{\partial x_{\beta}}\frac{\partial a^{\sigma}\left(  x\right)  }{\partial
x_{\gamma}}\frac{\partial a^{\rho}\left(  x\right)  }{\partial x_{\lambda}%
}\right]  .
\end{align*}

Defining $f_{\beta\tau}\left(  x\right)  $ as the double of the antisymetric
part of $\frac{\partial}{\partial x_{\beta}}a^{\tau}\left(  y\right)  $, i.e.,
$f^{\beta\tau}\left(  x\right)  =\frac{\partial}{\partial x_{\beta}}a^{\tau
}\left(  y\right)  -\frac{\partial}{\partial x_{\tau}}a^{\beta}\left(
y\right)  $, and realizing that the symmetric part of $\frac{\partial
}{\partial x_{\beta}}a^{\tau}\left(  y\right)  $ becomes zero after the
contraction with $\mathcal{P}_{\alpha\beta\gamma\lambda\mu\tau\sigma\rho}$
(because it is anti-symmetric under the interchange of the respective pair of
indexes: $\beta\leftrightarrow\tau$, $\gamma\leftrightarrow\sigma$ and
$\lambda\leftrightarrow\rho$), we get%
\[
j_{\mu}^{\mathrm{nl}}\left(  x\right)  =\frac{\mathcal{P}_{\alpha\beta
\gamma\lambda\mu\tau\sigma\rho}}{48}\frac{\partial}{\partial x_{\alpha}%
}\left[  f^{\beta\tau}\left(  x\right)  f^{\gamma\sigma}\left(  x\right)
f^{\lambda\rho}\left(  x\right)  \right]  .
\]

So the nonlinearly induced current is given by%
\begin{equation}
j_{\mu}^{\mathrm{nl}}\left(  x\right)  =\frac{1}{4}\mathfrak{L}_{\mathfrak{FF}%
}\frac{\partial}{\partial x_{\alpha}}\left[  f_{\alpha\mu}\left(  x\right)
f_{\beta\gamma}\left(  x\right)  f^{\beta\gamma}\left(  x\right)  \right]
+\frac{1}{4}\mathfrak{L}_{\mathfrak{GG}}\frac{\partial}{\partial x_{\alpha}%
}\left[  \tilde{f}_{\alpha\mu}\left(  x\right)  f_{\beta\gamma}\left(
x\right)  \tilde{f}^{\beta\gamma}\left(  x\right)  \right]  .
\label{General 4-current}%
\end{equation}

\emph{ }

In terms of the field strengths $\mathbf{E=E}\left(  x\right)  $ and
$\mathbf{B=B}\left(  x\right)  $, defined as $E_{i}\left(  x\right)
=f_{i0}\left(  x\right)  $, and $B_{i}\left(  x\right)  =\tilde{f}_{i0}\left(
x\right)  =\frac{1}{2}\epsilon_{ijk}f^{jk}\left(  x\right)  $, and bearing in
mind that $f_{\beta\gamma}\left(  x\right)  f^{\beta\gamma}\left(  x\right)
=2\left(  \mathbf{B}^{2}\left(  x\right)  -\mathbf{E}^{2}\left(  x\right)
\right)  $ and $f_{\beta\gamma}\left(  x\right)  \tilde{f}^{\beta\gamma
}\left(  x\right)  =-4\left(  \mathbf{E}\left(  x\right)  \cdot\mathbf{B}%
\left(  x\right)  \right)  ,$ we obtain%
\begin{equation}
j_{\mu}^{\mathrm{nl}}\left(  x\right)  =\frac{1}{2}\mathfrak{L}_{\mathfrak{FF}%
}\frac{\partial}{\partial x_{\alpha}}\left[  f_{\alpha\mu}\left(  x\right)
\left(  \mathbf{B}^{2}\left(  x\right)  -\mathbf{E}^{2}\left(  x\right)
\right)  \right]  -\mathfrak{L}_{\mathfrak{GG}}\frac{\partial}{\partial
x_{\alpha}}\left[  \tilde{f}_{\alpha\mu}\left(  x\right)  \left(
\mathbf{E}\left(  x\right)  \cdot\mathbf{B}\left(  x\right)  \right)  \right]
. \label{CAIO ANSWER 2: FIX SIGN ON L_GG}%
\end{equation}

Defining $j_{\mu}^{\mathrm{nl}}\left(  x\right)  =\left(  j_{0}^{\mathrm{nl}%
}\left(  \mathbf{r}\right)  ,-\mathbf{j}^{\mathrm{nl}}\left(  \mathbf{r}%
\right)  \right)  $, one can get the general nonlinear current in terms of the
electromagnetic field%
\begin{align}
&  j_{0}^{\mathrm{nl}}\left(  x\right)  =\frac{1}{2}\mathfrak{L}%
_{\mathfrak{FF}}\boldsymbol{\nabla}\cdot\left[  \left(  \mathbf{B}%
^{2}-\mathbf{E}^{2}\right)  \mathbf{E}\right]  -\mathfrak{L}_{\mathfrak{GG}%
}\boldsymbol{\nabla}\cdot\left[  \left(  \mathbf{E}\cdot\mathbf{B}\right)
\mathbf{B}\right]  ,\label{Explicit current}\\
&  \mathbf{j}^{\mathrm{nl}}\left(  x\right)  =-\frac{1}{2}\mathfrak{L}%
_{\mathfrak{FF}}\left(  \frac{\partial}{\partial t}\left[  \left(
\mathbf{B}^{2}-\mathbf{E}^{2}\right)  \mathbf{E}\right]  +\boldsymbol{\nabla
}\times\left[  \left(  \mathbf{B}^{2}-\mathbf{E}^{2}\right)  \mathbf{B}%
\right]  \right) \nonumber\\
&  +\mathfrak{L}_{\mathfrak{GG}}\left(  \frac{\partial}{\partial t}\left[
\left(  \mathbf{E}\cdot\mathbf{B}\right)  \right]  \mathbf{B}%
+\boldsymbol{\nabla}\times\left[  \left(  \mathbf{E}\cdot\mathbf{B}\right)
\mathbf{E}\right]  \right)  . \label{Explicit current2}%
\end{align}

\section{Stationary nonlinear Maxwell equations in blank vacuum}

\bigskip We want to solve (\ref{field equation}),
(\ref{Non linear induced current}) imposing time independence, so we define
the nonlinear electric field $\mathbf{E}^{\mathrm{nl}}$ and magnetic induction
$\mathbf{B}^{\mathrm{nl}}$ as%
\begin{align}
\mathbf{E}^{\mathrm{nl}}\left(  \mathbf{r}\right)   &  =\boldsymbol{\nabla
}a_{\text{\textrm{nl}}}^{0}(\mathbf{r}),\nonumber\\
\mathbf{B}^{\mathrm{nl}}\left(  \mathbf{r}\right)   &  =\boldsymbol{\nabla
}\times\boldsymbol{a}_{\text{\textrm{nl}}}(\mathbf{r}).
\label{nonlinear potentials}%
\end{align}

According to (\ref{nonlinfield}) these satisfy the following Maxwell equations
with the stationary source $j_{\mu}^{\mathrm{nl}}\left(  x\right)  $%
\begin{align}
\boldsymbol{\nabla}\cdot\mathbf{E}^{\mathrm{nl}}\left(  \mathbf{r}\right)   &
=j_{0}^{\mathrm{nl}}\left(  \mathbf{r}\right)  ,\label{Maxwell's Equation}\\
\boldsymbol{\nabla}\times\mathbf{E}^{\mathrm{nl}}\left(  \mathbf{r}\right)
&  =\mathbf{0},\nonumber\\
\boldsymbol{\nabla}\cdot\mathbf{B}^{\mathrm{nl}}\left(  \mathbf{r}\right)   &
=0,\nonumber\\
\boldsymbol{\nabla}\times\mathbf{B}^{\mathrm{nl}}\left(  \mathbf{r}\right)
&  =\mathbf{j}^{\mathrm{nl}}\left(  \mathbf{r}\right)  .\nonumber
\end{align}
Note, that owing to the absence of linear response (\ref{Linear response}),
field strengths and inductions are the same: $\mathbf{E}^{\mathrm{nl}}\left(
\mathbf{r}\right)  =\mathbf{D}^{\mathrm{nl}}\left(  \mathbf{r}\right)  ,$
$\mathbf{B}^{\mathrm{nl}}\left(  \mathbf{r}\right)  =\mathbf{H}^{\mathrm{nl}%
}\left(  \mathbf{r}\right)  .$

It is important to keep in mind that these Maxwell equations refer to HL units.

\subsection{Electrostatics}

To find the nonlinear electric field for a pure electrostatic problem, i.e.,
setting $\mathbf{B}\left(  \mathbf{r}\right)  =0$, we start from the nonlinear
charge density correction due to the fourth-rank tensor
(\ref{Explicit current}):
\begin{equation}
j_{0}^{\mathrm{nl}}\left(  \mathbf{r}\right)  =-\frac{1}{2}\mathfrak{L}%
_{\mathfrak{FF}}\boldsymbol{\nabla}\cdot\left[  \mathbf{E}\left(
\mathbf{r}\right)  E^{2}\left(  \mathbf{r}\right)  \right]  ,
\label{Nonlinear charge}%
\end{equation}
bearing in mind that the vector current density disappears in the
electrostatic case, $\mathbf{j}^{\mathrm{nl}}\left(  \mathbf{r}\right)  =0,$
according to (\ref{Explicit current2}). Using the first equation from
(\ref{Maxwell's Equation}), one can write
\begin{equation}
\boldsymbol{\nabla}\cdot\left[  \mathbf{E}^{\mathrm{nl}}\left(  \mathbf{r}%
\right)  +\frac{1}{2}\mathfrak{L}_{\mathfrak{FF}}\mathbf{E}\left(
\mathbf{r}\right)  E^{2}\left(  \mathbf{r}\right)  \right]  =0.
\label{equation}%
\end{equation}
We introduce the notation%
\begin{equation}
\boldsymbol{\mathcal{E}}\left(  \mathbf{r}\right)  =-\frac{1}{2}%
\mathfrak{L}_{\mathfrak{FF}}\mathbf{E}\left(  \mathbf{r}\right)  E^{2}\left(
\mathbf{r}\right)  . \label{Cubic electric field}%
\end{equation}

The general solution to equation (\ref{equation}) is defined up to the curl of
some vector field $\mathbf{\Omega}\left(  \mathbf{r}\right)  $:%
\begin{equation}
\mathbf{E}^{\mathrm{nl}}\left(  \mathbf{r}\right)  =\boldsymbol{\mathcal{E}%
}\left(  \mathbf{r}\right)  +\left[  \boldsymbol{\nabla}\times\mathbf{\Omega
}\left(  \mathbf{r}\right)  \right]  . \label{First equation for nonlinear E}%
\end{equation}

The second equation from (\ref{Maxwell's Equation}) allows one to fix
$\mathbf{\Omega}\left(  \mathbf{r}\right)  $ according to
\begin{equation}
\left[  \boldsymbol{\nabla}\times\left[  \boldsymbol{\nabla}\times
\mathbf{\Omega}\left(  \mathbf{r}\right)  \right]  \right]  =-\left[
\boldsymbol{\nabla}\times\boldsymbol{\mathcal{E}}\left(  \mathbf{r}\right)
\right]  , \label{omega}%
\end{equation}
in other words the vector field $\mathbf{\Omega}\left(  \mathbf{r}\right)  $
must satisfy the following Poison's equation
\begin{equation}
-\nabla^{2}\mathbf{\Omega}\left(  \mathbf{r}\right)  +\boldsymbol{\nabla
}\left(  \boldsymbol{\nabla}\cdot\mathbf{\Omega}\left(  \mathbf{r}\right)
\right)  =-\left[  \boldsymbol{\nabla}\times\boldsymbol{\mathcal{E}}\left(
\mathbf{r}\right)  \right]  . \label{after omega}%
\end{equation}
The field $\mathbf{\Omega}\left(  \mathbf{r}\right)  $ is defined by
(\ref{omega}) up to a gradient: the transformation $\mathbf{\Omega}\left(
\mathbf{r}\right)  \rightarrow\mathbf{\Omega}^{\prime}=\mathbf{\Omega
}+\boldsymbol{\nabla}\lambda$ leaves Eq. (\ref{omega}) intact. By choosing
$\lambda$ to satisfy the equation $\nabla^{2}\lambda=-\boldsymbol{\nabla}%
\cdot\mathbf{\Omega\left(  \mathbf{r}\right)  }$ we come to
$\boldsymbol{\nabla}\cdot\mathbf{\Omega}^{\prime}=0.$ Then the transformed
equation (\ref{after omega}) becomes (we omit the prime)%
\[
\nabla^{2}\mathbf{\Omega}\left(  \mathbf{r}\right)  =\left[
\boldsymbol{\nabla}\times\boldsymbol{\mathcal{E}}\left(  \mathbf{r}\right)
\right]  .
\]

Therefore, the solution (\ref{First equation for nonlinear E}) to the Maxwell
equation is the longitudinal projection of the field $\boldsymbol{\mathcal{E}%
}\left(  \mathbf{r}\right)  $ (\ref{Cubic electric field}), i.e.,%
\begin{equation}
E_{i}^{\mathrm{nl}}\left(  \mathbf{r}\right)  =\frac{\nabla_{i}\nabla_{j}%
}{\nabla^{2}}\mathcal{E}_{j}\left(  \mathbf{r}\right)  =-\frac{\nabla
_{i}\nabla_{j}}{4\pi}\int\frac{\mathcal{E}_{j}\left(  \mathbf{r}^{\prime
}\right)  }{\left\vert \mathbf{r-r}^{\prime}\right\vert }\mathrm{d}%
\mathbf{r}^{\prime}. \label{Solution in terms of projection operator}%
\end{equation}

or, equivalently,%
\begin{equation}
E_{i}^{\mathrm{nl}}\left(  \mathbf{r}\right)  =-\frac{\nabla_{i}}{4\pi}%
\int\frac{1}{\left\vert \mathbf{r-r}^{\prime}\right\vert }\nabla_{j}^{\prime
}\mathcal{E}_{j}\left(  \mathbf{r}^{\prime}\right)  \mathrm{d}\mathbf{r}%
^{\prime}. \label{Enl}%
\end{equation}

Note that the substitution of the Coulomb field of a point-like charge into
(\ref{Solution in terms of projection operator}) or into (\ref{Enl}) would
cause the divergency of the integral near $r^{\prime}=0:$ the present approach
fails near the point charge, since it is not applicable to its strongly
inhomogeneous field. Dealing with the point charge would require going beyond
the infrared approximation followed to in the present work. Nevertheless,
(\ref{Solution in terms of projection operator}) (or (\ref{Enl})) is sound as
applied to extended charges to be considered in the next section.

The nonlinear scalar potential corresponding to the field (\ref{Enl})
\begin{equation}
a_{\mathrm{nl}}^{0}\left(  r\right)  =\frac{1}{4\pi}\int\frac
{\boldsymbol{\nabla}^{\prime}\cdot\boldsymbol{\mathcal{E}}\left(
\mathbf{r}^{\prime}\right)  }{\left\vert \mathbf{r-r}^{\prime}\right\vert
}\mathrm{d}\mathbf{r}^{\prime}. \label{Scalar potential}%
\end{equation}

\subsection{Magnetostatics}

One can calculate the nonlinear magnetic field similarly to the previous
Subsection. Define%
\begin{equation}
\boldsymbol{\mathfrak{h}}\left(  \mathbf{r}\right)  =-\frac{1}{2}%
\mathfrak{L}_{\mathfrak{FF}}\mathbf{B}\left(  \mathbf{r}\right)  B^{2}\left(
\mathbf{r}\right)  . \label{Cubic magnetic field}%
\end{equation}
By setting $\mathbf{E}\left(  \mathbf{r}\right)  =0$ in
(\ref{Explicit current}), (\ref{Explicit current2}) we find the nonlinear
current for the case, where only magnetic field is developed:%

\begin{align}
&  j_{0}^{\mathrm{nl}}\left(  x\right)  =0,\\
&  \mathbf{j}^{\mathrm{nl}}\left(  x\right)  =\boldsymbol{\nabla}%
\times\boldsymbol{\mathfrak{h}}\left(  \mathbf{r}\right)  .\nonumber
\end{align}
Then it follows from the fourth equation in (\ref{Maxwell's Equation})) that%

\begin{equation}
\mathbf{B}^{\mathrm{nl}}\left(  \mathbf{r}\right)  =\boldsymbol{\mathfrak{h}%
}\left(  \mathbf{r}\right)  +\boldsymbol{\nabla}\Phi\left(  \mathbf{r}\right)
. \label{Magnetic projections}%
\end{equation}
Now by the third equation from (\ref{Maxwell's Equation}) one has
\[
\boldsymbol{\nabla}\cdot\mathbf{B}^{\mathrm{nl}}\left(  \mathbf{r}\right)
=\boldsymbol{\nabla}\cdot\boldsymbol{\mathfrak{h}}\left(  \mathbf{r}\right)
+\nabla^{2}\Phi\left(  \mathbf{r}\right)  =0.
\]
Therefore, as in the electric case, the scalar field $\Phi\left(
\mathbf{r}\right)  $ must satisfy the following Poisson equation
\[
\nabla^{2}\Phi\left(  \mathbf{r}\right)  =-\boldsymbol{\nabla}\cdot
\boldsymbol{\mathfrak{h}}\left(  \mathbf{r}\right)  .
\]
Hence, one can find the magnetic induction $\mathbf{B}^{\mathrm{nl}}\left(
\mathbf{r}\right)  $, bearing in mind (\ref{Magnetic projections}), as the
transverse projection of the field $\boldsymbol{\mathfrak{h}}\left(
\mathbf{r}\right)  $, i.e.,%
\begin{equation}
B_{i}^{\mathrm{nl}}\left(  \mathbf{r}\right)  =\left(  \delta_{ij}%
-\frac{\nabla_{i}\nabla_{j}}{\nabla^{2}}\right)  \mathfrak{h}_{j}\left(
\mathbf{r}\right)  =\mathfrak{h}_{i}\left(  \mathbf{r}\right)  +\frac
{\nabla_{i}\nabla_{j}}{4\pi}\int\frac{\mathfrak{h}_{j}\left(  \mathbf{r}%
^{\prime}\right)  }{\left\vert \mathbf{r-r}^{\prime}\right\vert }%
\mathrm{d}\mathbf{r}^{\prime}, \label{Bnl}%
\end{equation}
or%
\begin{equation}
B_{i}^{\mathrm{nl}}\left(  \mathbf{r}\right)  =\mathfrak{h}_{i}\left(
\mathbf{r}\right)  +\frac{1}{4\pi}\nabla_{i}\int\frac{\nabla_{j}^{\prime}%
\cdot\mathfrak{h}_{j}\left(  \mathbf{r}^{\prime}\right)  }{\left\vert
\mathbf{r-r}^{\prime}\right\vert }\mathrm{d}\mathbf{r}^{\prime}.
\label{Vector form for magnetic field}%
\end{equation}

The corresponding vector potential is%
\begin{equation}
\boldsymbol{a}_{\mathrm{nl}}\left(  \mathbf{r}\right)  =\frac{1}{4\pi}%
\int\frac{\boldsymbol{\nabla}^{\prime}\times\boldsymbol{\mathfrak{h}}\left(
\mathbf{r}^{\prime}\right)  }{\left\vert \mathbf{r-r}^{\prime}\right\vert
}\mathrm{d}\mathbf{r}^{\prime}. \label{Vector potential}%
\end{equation}

As far as practical applications of the equations derived in this Section for
the nonlinear electric (\ref{Enl}) and magnetic (\ref{Bnl}) fields are
concerned we shall as a matter of fact take their right-hand sides with the
linear approximations $\boldsymbol{E\simeq E}^{\text{lin}}\left(
\mathbf{r}\right)  =\boldsymbol{\nabla}a_{\text{\textrm{lin}}}^{0}%
(\mathbf{r}),$ $\boldsymbol{B\simeq B}^{\text{lin}}(\mathbf{r}%
)=\boldsymbol{\nabla}\times\boldsymbol{a}_{\text{\textrm{lin}}}(\mathbf{r})$
substituted into the expressions (\ref{Cubic electric field}) for
$\mathcal{E}$($\boldsymbol{r}$) and (\ref{Cubic magnetic field}) for
$\boldsymbol{\mathfrak{h}}\left(  \mathbf{r}\right)  ,$ respectively$.$
Nevertheless, certain issues can and will be discussed based on exactly
nonlinear equations. For this reason we shall, for a time being, retain exact
fields in (\ref{Solution in terms of projection operator}) and
(\ref{Vector form for magnetic field}).

\section{Applications to special static and stationary sources}

In this chapter, we are concerned about applications, like the nonlinear
correction to the fields due to a charged sphere, infinite thread, infinite
plane, infinite wire with a current, electric and magnetic dipoles.

\subsection{Sources providing degeneracy of the projection operators}

Equations (\ref{Solution in terms of projection operator}), (\ref{Bnl})
obtained above map the time-independent electric and magnetic fields to their
nonlinear counterparts. Generally, these mappings are nonlocal thanks to the
presence of the inverse differential operator $\nabla^{-2}.$ This fact also
makes them more difficult to calculate. However, there are some trivial cases
when a high symmetry of the problem turns the projection operators in
(\ref{Solution in terms of projection operator}) or (\ref{Bnl}) into unit
operators and thereby reduces the nonlinearity to a local form. In this
Subsection we concentrate on such cases, first of all on the spherically
symmetric case, which is very important, because it leads to the known
nonlinear correction to the Coulomb field, produced by an O(3)-symmetrically
charged sphere. If the radius of the charge is not too small (it must have low
derivative, as our assumption), it may describe the electric field produced by
the nuclei of some atoms. Furthermore, there are some cases, whose symmetry
includes translation invariance, where charge and current distributions are
not localized (infinite plane and infinite thread and wire), and it is easy to
show that the projection operator is the identity operator as well.

\subsubsection{Spherical symmetry: charged sphere}

Spherical symmetry implies that any vector may be directed only along the
radius-vector $\mathbf{r.}$ As applied to the vector
(\ref{Cubic electric field}) \ this rule reads: $\mathcal{E}_{i}\left(
\mathbf{r}\right)  =x_{i}\Lambda(r)$, where the scalar $\Lambda$ may depend
only on the distance $r=\left(  x_{k}^{2}\right)  ^{\frac{1}{2}}.$ It is easy
to check by direct differentiation that%
\begin{equation}
\nabla_{i}\nabla_{j}\left(  x_{j}\Lambda(r)\right)  =\text{ }\nabla^{2}%
x_{i}\Lambda(r), \label{Identity operator}%
\end{equation}
i.e., the projection acts as the identity operator: $\frac{\mathbf{\nabla}%
_{i}\mathbf{\nabla}_{j}}{\nabla^{2}}=\delta_{ij}$.

Then Eq.(\ref{Solution in terms of projection operator}) states that the
corrected nonlinear electric field in the entire space is just $\mathbf{E}%
^{\text{nl}}\left(  \mathbf{r}\right)  =\boldsymbol{\mathcal{E}}\left(
\mathbf{r}\right)  =-\frac{1}{2}\mathfrak{L}_{\mathfrak{FF}}E^{2}\left(
r\right)  \mathbf{E}\left(  \mathbf{r}\right)  $. Hence, within the cubic
approximation (\ref{General current}) the total (nonlinearly corrected)
electric field, equal to $\mathbf{E}\left(  \mathbf{r}\right)  =\mathbf{E}%
^{\text{lin}}\left(  \mathbf{r}\right)  +\mathbf{E}^{\text{nl}}\left(
\mathbf{r}\right)  $ in accord with (\ref{linear+nonlinear}), is subject to
the following cubic equation%
\begin{equation}
\mathbf{E}\left(  \mathbf{r}\right)  =\mathbf{E}^{\text{lin}}\left(
\mathbf{r}\right)  -\frac{1}{2}\mathfrak{L}_{\mathfrak{FF}}\mathbf{E}\left(
\mathbf{r}\right)  E^{2}\left(  r\right)  .
\label{Symmetric nonlinear correction}%
\end{equation}
With the Euler-Heisenberg Lagrangian \cite{BerLifPit} one can calculate that
$\mathfrak{L}_{\mathfrak{FF}}=\frac{e^{4}}{45\pi^{2}m^{4}},$ where $m$ is the
electron mass, while its charge $e$ is to be taken in the same system of units
as $\mathfrak{L}_{\mathfrak{FF}}$.

As long as this coefficient is very small it hardly makes sense to treat the
nonlinearity in equation (\ref{Symmetric nonlinear correction}) seriously.
Instead, it is sufficient to keep $\mathbf{E}\left(  \mathbf{r}\right)
\simeq\mathbf{E}^{\text{lin}}\left(  \mathbf{r}\right)  $ in its right-hand
side. Then the approximate solution to this equation is given as%
\begin{equation}
\mathbf{E}\left(  \mathbf{r}\right)  =\mathbf{E}^{\text{lin}}\left(
\mathbf{r}\right)  \left(  1-\frac{2\alpha}{45\pi}\left(  \frac{eE^{\text{lin}%
}\left(  r\right)  }{m^{2}}\right)  ^{2}\right)  . \label{Euler-Heisenberg}%
\end{equation}
This expression can be trusted for the field values $E^{\text{lin}}\left(
\mathbf{r}\right)  \lesssim\frac{m^{2}}{e}\sqrt{\frac{45\pi}{2\alpha}}$. This
upper bound of applicability is higher than Schwinger's characteristic value
$E_{\text{Sch}}=$\ $\frac{m^{2}}{e}.$ However, if the field compares with this
characteristic value and exceeds it, the vacuum instability via spontaneous
production of electron-positron pairs \cite{FGS} is expected to become more
and more essential, which fact would require a revision of our result.

If the standard Coulomb field $\frac{q}{4\pi r^{2}}\frac{\mathbf{r}}{r}$
linearly produced by a point charge $q$ is used for $\mathbf{E}^{\text{lin}%
}\left(  \mathbf{r}\right)  $ in (\ref{Euler-Heisenberg}) the latter is
reduced to an expression for the nonlinear correction to this field in the
vacuum found in \cite{BerLifPit} , which is highly singular in the origin
$r=0$. For the atomic field with $q=Ze$ and $Z$\ about a few tens, the
essential correction of $10^{-5}$ is achieved at the distance from the nucleus
about the electron Compton length.

Equation (\ref{Euler-Heisenberg}) can be applied to an extended
spherically-symmetric charge $q$, say, a homogeneously charged sphere with the
radius $R$. We define two regions $r<R$ and $r>R$, and the linear electric
field is given by
\begin{equation}
\mathbf{E}^{\text{lin}}\left(  r\right)  =\Theta\left(  R-r\right)
\mathbf{E}^{<}\left(  r\right)  +\Theta\left(  r-R\right)  \mathbf{E}%
^{>}\left(  r\right)  , \label{sphere}%
\end{equation}
where
\begin{equation}
\mathbf{E}^{<}\left(  r\right)  =\frac{qr}{4\pi R^{3}}\mathbf{e}_{r}%
\mathbf{,}\text{ \ }\mathbf{E}^{>}\left(  r\right)  =\frac{q}{4\pi r^{2}%
}\mathbf{e}_{r}\mathbf{.} \label{sphere <}%
\end{equation}

This field should be used in ((\ref{Euler-Heisenberg}) to produce a finite
value in the origin $r=0.$

\subsubsection{Cylindric plus translation symmetry: charged plane, charged
thread, and current-carrying wire}

Three other cases, when the projections in
(\ref{Solution in terms of projection operator})\ or in (\ref{Bnl}) reduce to
trivial identities may be revealed.

One of them is supplied by the translation invariance along every plane
orthogonal to a given axis (axis 3, for instance), characteristic of a
finite-thickness infinite plane charged homogeneously along the directions 1
and 2. Then $\mathcal{E}_{1,2}\left(  \mathbf{r}\right)  =$ 0, $\mathcal{E}%
_{3}\left(  \mathbf{r}\right)  =x_{3}\Lambda(x_{3}),$ hence, when applied to
$\boldsymbol{\mathcal{E}}\left(  \mathbf{r}\right)  ,$ the projection operator
is unity:$\frac{\nabla_{i}\nabla_{j}}{\nabla^{2}}=\delta_{ij}.$ Then, in place
of ((\ref{Euler-Heisenberg}) we get%

\begin{equation}
E_{3}\left(  x_{3}\right)  =E_{3}^{\text{lin}}\left(  x_{3}\right)  \left(
1-\frac{2\alpha}{45\pi}\left(  \frac{eE^{\text{lin}}\left(  x_{3}\right)
}{m^{2}}\right)  ^{2}\right)  ,\text{ \ }E_{1,2}=0. \label{46a}%
\end{equation}
If, besides, the volume charge density $\kappa$ is also homogeneous along the
axis 3, so that finally $\kappa=const,$ and the \ plane is cituated
symmetrically with respect to the coordinate plane $x_{3}=0,$ its standard
electric field
\begin{equation}
\mathbf{E}^{\text{lin}}\left(  x_{3}\right)  =\Theta\left(  \frac{d}%
{2}-\left\vert x_{3}\right\vert \right)  \mathbf{E}^{<}\left(  x_{3}\right)
+\Theta\left(  \left\vert x_{3}\right\vert -\frac{d}{2}\right)  \mathbf{E}%
^{>}\left(  x_{3}\right)  , \label{plan}%
\end{equation}%
\begin{equation}
\mathbf{E}^{<}\left(  x_{3}\right)  =\kappa x_{3}\mathbf{e}_{3}\mathbf{,}%
\text{ \ }\mathbf{E}^{>}\left(  x_{3}\right)  =\frac{\kappa d}{2}%
\mathrm{sgn}\left(  x_{3}\right)  \mathbf{e}_{3}\mathbf{,} \label{plane <}%
\end{equation}
where $\mathrm{sgn}\left(  x_{3}\right)  =\frac{\left\vert x_{3}\right\vert
}{x_{3}}$ and $d$ is the thickness of the plane, should be used in (\ref{46a})
for obtaining the nonlinear extension of the field produced by this charged plane.

Another case of trivialization of the projection operator is provided by the
cylindric symmetry under rotations around the axis 3 supplemented by the
invariance under translations along this axis. This type of symmetry is
peculiar to an infinite round-cylindric thread homogeneously charged along
itself and O(2)-symmetrically charged across. Then $\mathcal{E}_{3}\left(
\mathbf{r}\right)  =0,$ $\mathcal{E}_{1,2}\left(  \mathbf{r}\right)
=\rho_{1,2}\Lambda(\rho),$ where $\boldsymbol{\rho}$ \ is the radius-vector in
the (1,2)-plane. In this case, it is again easy to check by direct
differentiation that $\boldsymbol{\nabla}(\boldsymbol{\nabla\cdot\rho}%
\Lambda(\rho))=\nabla^{2}\boldsymbol{\rho}\Lambda(\rho)$ and hence the
projection operator in (\ref{Solution in terms of projection operator}) is
unity. Then, in place of (\ref{Euler-Heisenberg}) the relation

\bigskip%
\begin{equation}
E_{1,2}\left(  \rho\right)  =E_{1,2}^{\text{lin}}\left(  \rho\right)  \left(
1-\frac{2\alpha}{45\pi}\left(  \frac{e|\mathbf{E}^{\text{lin}}\left(
\rho\right)  |}{m^{2}}\right)  ^{2}\right)  ,\text{ \ }E_{3}=0. \label{46b}%
\end{equation}
should hold. If, besides, the volume charge density $\kappa$ is independent
also on the radius $\rho,$ $\kappa=const,$ the standard electric field of the
homogeneously charged straight thread with the radius $R$ and its axis
coinciding with the coordinate axis 3%

\begin{equation}
\mathbf{E}^{\text{lin}}\left(  \rho\right)  =\Theta\left(  R-\rho\right)
\mathbf{E}^{<}\left(  \rho\right)  +\Theta\left(  \rho-R\right)
\mathbf{E}^{>}\left(  \rho\right)  ,
\end{equation}
\ \
\begin{equation}
\mathbf{E}^{<}\left(  \rho\right)  =\frac{\kappa\rho}{2}\mathbf{e}_{\rho
}\mathbf{,}\text{ }\mathbf{E}^{>}\left(  \rho\right)  =\frac{\kappa R^{2}%
}{2\rho}\mathbf{e}_{\rho},\text{ }\mathbf{e}_{\rho}=\frac{\boldsymbol{\rho}%
}{\rho}%
\end{equation}
is to be substituted into (\ref{46b}).

The same symmetry as that of the infinite thread above is inherent to the
straight infinite round-cylindric wire carrying a

constant current with O(2)-invariant density. This time, the field
(\ref{Cubic magnetic field}) has the structure $\boldsymbol{\mathfrak{h}%
}\left(  \mathbf{r}\right)  =\boldsymbol{e}_{3}\times\mathbf{r}$
$\Lambda\left(  |\boldsymbol{e}_{3}\times\mathbf{r|}\right)  $, where
$\boldsymbol{e}_{3}$ is the unit vector along axis 3, and $|\boldsymbol{e}%
_{3}\times\mathbf{r|=}\rho.$ Now $\boldsymbol{\nabla\cdot\mathfrak{h}}\left(
\mathbf{r}\right)  =0,$ and the projection operator in (\ref{Bnl}) is unity
again, so that $B_{i}^{\mathrm{nl}}\left(  \mathbf{r}\right)  =\mathfrak{h}%
_{i}\left(  \mathbf{r}\right)  .$

Now if we take a fully homogeneous current density $\mathbf{j}$ per unit area,
flowing along the positive direction, and calculate the linear magnetic field
produced by it, we get the standard relation
\begin{equation}
\mathbf{B}^{\text{lin}}\left(  \rho\right)  =\Theta\left(  R-\rho\right)
\mathbf{B}^{<}\left(  \rho\right)  +\Theta\left(  \rho-R\right)
\mathbf{B}^{>}\left(  \rho\right)  , \label{thread}%
\end{equation}
\ \
\begin{equation}
\mathbf{B}^{<}\left(  \rho\right)  =\frac{j\rho}{2}\mathbf{e}_{\phi}%
\mathbf{,}\text{ }\mathbf{B}^{>}\left(  \rho\right)  =\frac{jR^{2}}{2\rho
}\mathbf{e}_{\phi},\text{ }\mathbf{e}_{\phi}=\frac{\mathbf{e}_{3}%
\times\boldsymbol{\rho}}{\rho} \label{thread 1}%
\end{equation}
to be used in the expression%

\begin{equation}
B_{\phi}\left(  \rho\right)  =B_{\phi}^{\text{lin}}\left(  \rho\right)
\left(  1-\frac{2\alpha}{45\pi}\left(  \frac{eB_{\phi}^{\text{lin}}\left(
\rho\right)  }{m^{2}}\right)  ^{2}\right)  \label{Euler-Heisenberg2}%
\end{equation}
that follows from (\ref{Bnl}) and (\ref{Cubic magnetic field}) for the present
type of symmetry of the problem.

\subsection{Cylindric symmetry: Elementary dipoles}

In this Subsection, we proceed with other cases of cylindric symmetry , but
without any translation invariance. We will calculate nonlinear corrections to
the fields of magnetic and electric dipoles by realizing nontrivial
projections in (\ref{Solution in terms of projection operator})\ or in
(\ref{Bnl}) .

\subsubsection{Magnetic dipole.}

Let there be a sphere with the radius $R,$ and a time-independent current
$\mathbf{j}(\mathbf{r})$ concentrated on its surface:%
\begin{equation}
\mathbf{j}(\mathbf{r})=3\frac{\boldsymbol{\mathcal{M}}\times\mathbf{r}}{r^{4}%
}\delta(r-R). \label{j}%
\end{equation}
Here $\boldsymbol{\mathcal{M}}$ is a constant vector directed, say, along the
axis 3. The current density (\ref{j}) obeys the continuity condition
$\nabla\cdot\mathbf{j}(\mathbf{r})$ $=0$, its flow lines are circular in the
planes parallel to the plane (1,2). The magnetic field produced by this
current via the Maxwell equation $\nabla\times\boldsymbol{B}^{\text{lin}%
}(\mathbf{r})\boldsymbol{=}$ $\mathbf{j}(\mathbf{r})$ is, everywhere outside
of the sphere, the magnetic dipole field \footnote{The extra factor of 4$\pi$
should appear in (\ref{j}) in the Gauss system in accordance with the fact
that this factor is present in the right-hand side of the Maxwell equation
used.}%
\begin{equation}
\mathbf{B}^{>}=-\frac{\boldsymbol{\mathcal{M}}}{r^{3}}+3\frac{\mathbf{r}\text{
}\cdot\boldsymbol{\mathcal{M}}}{r^{5}}\mathbf{r}\boldsymbol{,} \label{magdip>}%
\end{equation}
with the constant vector density $\boldsymbol{\mathcal{M}}$ introduced in
(\ref{j}) playing the role of the corresponding magnetic moment. Inside the
sphere the magnetic field is constant:
\begin{equation}
\mathbf{B}^{<}=\frac{2\boldsymbol{\mathcal{M}}}{R^{3}}.\text{ }
\label{magdip<}%
\end{equation}
It turns to infinity for a point-like dipole $R=0.$\ 

We should stress that Eqs. (\ref{magdip>}), (\ref{magdip<}) are used to difine
the magnetic moment both in the (rationalized) Heaviside-Lorentz and Gauss
systems of units, provided that the right- and left-hand sides refer to one
and the same system. This form of equation may be called dimension-covariant.

We have
\begin{equation}
\mathbf{B}^{\text{lin}}(\mathbf{r})=\text{ }\Theta\left(  R-r\right)
\mathbf{B}^{<}\left(  \mathbf{r}\right)  +\Theta\left(  r-R\right)
\mathbf{B}^{>}\left(  \mathbf{r}\right)  , \label{Blin}%
\end{equation}
where $\Theta\left(  x\right)  =\left\{
\begin{array}
[c]{c}%
1,\text{ \ if }x>0,\\
0,\text{ \ \ if }x<0.
\end{array}
\right.  $is the step function. Each function $\mathbf{B}^{\gtrless}\left(
\mathbf{r}\right)  $ satisfies the sourceless Maxwell equation $\nabla
\times\boldsymbol{B}^{\gtrless}\left(  \boldsymbol{r}\right)  =0,$ whereas the
delta function in the current density (\ref{j}) is produced by differentiation
of the step function in (\ref{Blin}) under the curl operation. The relative
coefficient 2 in (\ref{magdip<}) is chosen so as to make the field also
satisfy the other Maxwell equation $\nabla\cdot\boldsymbol{B}^{\text{lin}%
}\left(  \boldsymbol{r}\right)  =0.$ Consequently, the radial component
$\boldsymbol{r}\cdot\mathbf{B}^{\text{lin}}(\mathbf{r})$\ of the field
(\ref{Blin}) is continuous at the border of the sphere $r=R,$ while its
tangent component in the plane spanned by the vectors $\boldsymbol{r}$ and
$\boldsymbol{\mathcal{M}},$ is not. (Its component in the plane orthogonal to
$\boldsymbol{r}$ and $\boldsymbol{\mathcal{M}}$ disappears). We write the
field (\ref{Blin}) as
\begin{align}
&  \mathbf{B}^{\text{lin}}\left(  \mathbf{r}\right)  =\left(  \frac
{2\mathcal{M}}{R^{3}}\cos\theta\mathbf{e}_{r}-\frac{2\mathcal{M}}{R^{3}}%
\sin\theta\mathbf{e}_{\theta}\right)  \Theta\left(  R-r\right) \nonumber\\
&  +\left(  \frac{2\mathcal{M}}{r^{3}}\cos\theta\mathbf{e}_{r}+\frac
{\mathcal{M}}{r^{3}}\sin\theta\mathbf{e}_{\theta}\right)  \Theta\left(
r-R\right)  , \label{Linear magnetic field1}%
\end{align}
where $\theta$ is the angle between the vectors $\boldsymbol{r}$ and
$\boldsymbol{\mathcal{M}},$ the unit vectors $\mathbf{e}_{r}$ and
$\mathbf{e}_{\theta}$ are, respectively, along $\boldsymbol{r}$ and along the
tangent direction in the plane spanned by these vectors, directed in such a
way that $\mathbf{e}_{\theta}$ be opposite to $\boldsymbol{\mathcal{M}}$
at\emph{ }$\theta=\pi/2$ \emph{. }Let us calculate the field
(\ref{Cubic magnetic field}) taken on the linear expression
(\ref{Linear magnetic field1}) for $\boldsymbol{B}$ \
\begin{align}
&  -\frac{2\boldsymbol{\mathfrak{h}}\left(  \mathbf{r}\right)  R^{9}%
}{\mathfrak{L}_{\mathfrak{FF}}\mathcal{M}^{3}}=\left(  8\cos\theta
\mathbf{e}_{r}-8\sin\theta\mathbf{e}_{\theta}\right)  \Theta\left(  R-r\right)
\nonumber\\
&  +\left[  \left(  2\frac{R^{9}}{r^{9}}\cos\theta+6\frac{R^{9}}{r^{9}}%
\cos^{3}\theta\right)  \mathbf{e}_{r}+\left(  4\frac{R^{9}}{r^{9}}\sin
\theta-3\frac{R^{9}}{r^{9}}\sin^{3}\theta\right)  \mathbf{e}_{\theta}\right]
\Theta\left(  r-R\right)  . \label{cubic}%
\end{align}
The further calculations of the nonlinear magnetic field of a dipole following
(\ref{Vector form for magnetic field}) are traced in Appendix 2. The result
is
\begin{align}
&  \frac{-2R^{9}}{\mathfrak{L}_{\mathfrak{FF}}\mathcal{M}^{3}}\mathbf{B}%
^{\mathrm{nl}}\left(  \mathbf{r}\right)
=\label{Nonlinear magnetic field correction with full terms}\\
&  \left\{  \left[  \left(  \frac{88}{15}-\frac{324}{385}\frac{r^{2}}{R^{2}%
}\right)  \cos\theta+\frac{108}{77}\frac{r^{2}}{R^{2}}\cos^{3}\theta\right]
\mathbf{e}_{r}\right.  \left.  +\left[  \left(  -\frac{88}{15}-\frac{432}%
{385}\frac{r^{2}}{R^{2}}\right)  \sin\theta+\frac{108}{77}\frac{r^{2}}{R^{2}%
}\sin^{3}\theta\right]  \mathbf{e}_{\theta}\right\}  \Theta\left(  R-r\right)
\nonumber\\
&  +\left\{  \left[  \left(  \frac{28}{5}\frac{R^{3}}{r^{3}}-\frac{18}%
{35}\frac{R^{5}}{r^{5}}\allowbreak-\frac{2}{33}\frac{R^{9}}{r^{9}}\right)
\cos\theta+\left(  \frac{6}{7}\frac{R^{5}}{r^{5}}+\frac{6}{11}\frac{R^{9}%
}{r^{9}}\right)  \cos^{3}\theta\right]  \mathbf{e}_{r}\right. \nonumber\\
&  \left.  +\left[  \left(  \frac{14}{5}\frac{R^{3}}{r^{3}}+\frac{18}{35}%
\frac{R^{5}}{r^{5}}+\frac{56}{33}\frac{R^{9}}{r^{9}}\right)  \sin
\theta+\left(  -\frac{9}{14}\frac{R^{5}}{r^{5}}\allowbreak-\frac{21}{22}%
\frac{R^{9}}{r^{9}}\right)  \sin^{3}\theta\right]  \mathbf{e}_{\theta
}\right\}  \Theta\left(  r-R\right)  .\nonumber
\end{align}

For the inside region of the sphere, i.e., when $r<R$, we see that the
solution is regular at the origin and directed, as expected, along
$\boldsymbol{\mathcal{M}}$: $\mathbf{B}^{\mathrm{nl}}\left(  0\right)
=-\allowbreak\frac{44}{15R^{9}}\mathfrak{L}_{\mathfrak{FF}}\mathcal{M}%
^{3}\mathbf{e}_{3}$. For the outside region of the sphere, i.e., when $r>R$,
we see that the field has three powers of the ratio of the distance from the
origin to the radius of the sphere: the 9th, the 5th and, most interesting,
the 3rd power, the latter just like the linear field (\ref{Blin}). This means
that, at large distances from the origin, the correction to the field behaves
in the same way as the linear field. (This is in contrast to the field of
electric monopole given by (\ref{Euler-Heisenberg}) after the Coulimb field is
substituted into it, where the correction decreases as the 6th power, while
the linear (Coulomb) field decreases as the 2nd power.) Therefore, imposing
$r>>R$%
\begin{equation}
\left.  \mathbf{B}^{\mathrm{nl}}\left(  \mathbf{r}\right)  \right\vert
_{r>>R}=-\allowbreak\frac{7}{5}\mathfrak{L}_{\mathfrak{FF}}\frac
{\mathcal{M}^{3}}{R^{6}r^{3}}\left(  2\cos\theta\mathbf{e}_{r}+\sin
\theta\mathbf{e}_{\theta}\right)  . \label{>>}%
\end{equation}

In terms of the linear field $(\ref{Blin}$) the total (nonlinearly corrected)
field is the sum of the linear field and the nonlinear field, according to
(\ref{linear+nonlinear})
\begin{equation}
\left.  \mathbf{B}\left(  \mathbf{r}\right)  \right\vert _{r>>R}%
=\mathbf{B}^{\text{lin}}\left(  \mathbf{r}\right)  \left(  1-\frac{7}%
{5}\mathfrak{L}_{\mathfrak{FF}}\frac{\mathcal{M}^{2}}{R^{6}}\right)  .
\label{Final correction}%
\end{equation}
Note that unlike Eq. (\ref{Linear magnetic field1}), the expressions
(\ref{Nonlinear magnetic field correction with full terms}),\ (\ref{>>}),
(\ref{Final correction}) refer only to HL system, since Eq.
(\ref{Vector form for magnetic field}) is a solution to the Maxwell equation
(\ref{Maxwell's Equation}) written in that system.

For the Euler-Heisenberg Lagrangian one has%
\begin{equation}
\mathfrak{L}_{\mathfrak{FF}}=\frac{e^{4}}{45\pi^{2}m^{4}}=\frac{e^{2}}%
{45\pi^{2}}\left(  \frac{1}{B_{\text{Sch}}}\right)  ^{2}. \label{LFF}%
\end{equation}
Once the Euler-Heisenberg Lagrangian $\mathfrak{L}$ is a function of the
product $eB$ (or of $e^{2}\mathfrak{F),}$ it is dimension-invariant, i.e. the
same in HL and Gauss systems. Hence the variational derivative $\mathfrak{L}%
_{\mathfrak{FF}}$ (\ref{LFF}) is dimension-covariant. This implies that in it
one may take $e$ either equal to its Gauss value $e^{\text{G}}=(1/137)^{1/2}$
and, correspondingly, choose $B_{\text{Sch}}=B_{\text{Sch}}^{\text{G}}%
=m^{2}/e^{\text{G}}=4.4\cdot10^{13}$G, or, alternately, take $e=$
$e^{\text{HL}}=(4\pi/137)$ and $B_{\text{Sch}}=B_{\text{Sch}}^{\text{HL}%
}=m^{2}/e^{\text{Hl}}=1.24\cdot10^{13}$G. It is understood that in the two
cases the field squared $\mathfrak{F}$ inside $\mathfrak{L}_{\mathfrak{FF}}$
should be simultaneously normalized differently. However, when substituting
(\ref{LFF}) in (\ref{Final correction}) only HL values should be used. \ So,
in QED the overall field of the magnetic dipole is given by%
\begin{align}
&  \left.  \mathbf{B}\left(  \mathbf{r}\right)  \right\vert _{r>>R}%
=\mathbf{B}^{\text{lin}}\left(  \mathbf{r}\right)  \left(  1-\frac{7}{5}%
\frac{e^{2}}{45\pi^{2}}\left(  \frac{e}{m^{2}}\frac{\mathcal{M}}{R^{3}%
}\right)  ^{2}\right) \label{CAIO ANSWER 4: CORRECT DIMENSIONS}\\
&  =\mathbf{B}^{\text{lin}}\left(  \mathbf{r}\right)  \left(  1-\frac{7}%
{5}\frac{e^{2}}{180\pi^{2}}\left(  \frac{B^{\text{%
$<$%
}}}{B^{\text{Sch}}}\right)  ^{2}\right)  .\nonumber
\end{align}

\bigskip The last equation is to be presented in the final form, independent
of a system of units%
\begin{equation}
\left.  \mathbf{B}\left(  \mathbf{r}\right)  \right\vert _{r>>R}%
=\mathbf{B}^{\text{lin}}\left(  \mathbf{r}\right)  \left(  1-\frac{7}{5}%
\frac{\alpha}{45\pi}\left(  \frac{B^{\text{%
$<$%
}}}{B^{\text{Sch}}}\right)  ^{2}\right)  . \label{Final field}%
\end{equation}

\bigskip

Once the form (\ref{>>}) -- (\ref{Final field} ) of the magnetic dipole field
far from the source proved to be invariant under nonlinear correction, we may
abandon the approximation $\mathbf{B}\simeq\mathbf{B}^{\text{lin}}%
(\mathbf{r})$ used when calculating $\boldsymbol{\mathfrak{h}}\left(
\mathbf{r}\right)  $ above. This means that $\mathcal{M}$ in (\ref{cubic}),
(\ref{Nonlinear magnetic field correction with full terms}) and (\ref{>>}) may
be thought of as the final, nonlinearly corrected, magnetic moment
$\mathcal{M}^{\text{nlc}}$. Then it becomes subject to the cubic equation%
\begin{equation}
\mathcal{\boldsymbol{\mathcal{M}}}^{\text{nlc}}%
=\mathcal{\boldsymbol{\mathcal{M}}}-\frac{7}{5}\mathfrak{L}_{\mathfrak{FF}%
}\mathcal{\boldsymbol{\mathcal{M}}}^{\text{nlc}}\left(  \frac{\mathcal{M}%
^{\text{nlc}}}{R^{3}}\right)  ^{2}. \label{self-coupling}%
\end{equation}
This equation shows nonlinear self-coupling of the magnetic moment. The minus
sign here excludes its spontaneous production when the "bare" magnetic moment
is not present $\left(  \mathcal{M=}\text{ 0}\right)  $ as long as
$\mathfrak{L}_{\mathfrak{FF}}>0.$ (The positivity of $\mathfrak{L}%
_{\mathfrak{FF}}$ is a consequence of causality and unitarity \cite{Shabus}).
In the approximation of $\mathbf{B}\simeq\mathbf{B}^{\text{lin}}(\mathbf{r})$
the correction to the magnetic moment looks as
\begin{equation}
\mathcal{\boldsymbol{\mathcal{M}}}^{\mathrm{nlc}}%
=\mathcal{\boldsymbol{\mathcal{M}}}\left(  1-\frac{7}{5}\mathfrak{L}%
_{\mathfrak{FF}}\frac{\mathcal{M}^{2}}{R^{6}}\right)  . \label{Mnlc}%
\end{equation}

\subsubsection{Electric dipole.}

Let there be a sphere with the radius $R$, and a time-independent charge
$j_{0}\left(  \mathbf{r}\right)  $ concentrated on its surface:%
\begin{equation}
j_{0}\left(  \mathbf{r}\right)  =3\frac{\mathbf{p\cdot r}}{r^{4}}\delta\left(
r-R\right)  \label{1}%
\end{equation}

Here $\mathbf{p}$ is a constant vector directed, say, "to the North", i.e.
along the axis 3. The current density (\ref{1}) obeys the continuity condition
$\partial_{0}j_{0}=0$. It is continuously distributed over the surface, the
"northern" hemisphere being positevely, and "southern" hemisphere negatively
charged. The extremum charge densities are achieved at the poles, while the
equator remains neutral. The electric field produced by this current via the
Maxwell equation $\nabla\cdot\mathbf{E}^{\mathrm{lin}}\left(  \mathbf{r}%
\right)  =$ $j_{0}\left(  \mathbf{r}\right)  $ is, outside of the sphere
$r>R$, where $j_{0}\left(  \mathbf{r}\right)  =0,$ the electric dipole field%
\begin{equation}
\mathbf{E}^{>}\left(  \mathbf{r}\right)  =-\frac{\mathbf{p}}{r^{3}}%
+3\frac{\mathbf{r\cdot p}}{r^{5}}\mathbf{r}, \label{64}%
\end{equation}
with the constant vector density $\mathbf{p}$ playing the role of the
corresponding electric moment. Inside the sphere the electric field is
constant:%
\begin{equation}
\mathbf{E}^{<}\left(  \mathbf{r}\right)  =-\frac{\mathbf{p}}{R^{3}}.
\label{??}%
\end{equation}

It turns to infinity for a point-like dipole $R=0$. We have%
\begin{equation}
\mathbf{E}^{\mathrm{lin}}\left(  \mathbf{r}\right)  =\Theta\left(  R-r\right)
\mathbf{E}^{<}\left(  \mathbf{r}\right)  +\Theta\left(  r-R\right)
\mathbf{E}^{>}\left(  \mathbf{r}\right)  . \label{4}%
\end{equation}
Each function $\mathbf{E}^{\lessgtr}\left(  \mathbf{r}\right)  $ satisfies the
sourceless Maxwell equation $\nabla\cdot\mathbf{E}^{\lessgtr}\left(
\mathbf{r}\right)  =0$, whereas the delta function in the current density
(\ref{1}) is produced by differentiation of the step function $\Theta$ in
(\ref{4}) under the divergence operation in the Maxwell equation $\nabla
\cdot\mathbf{E}^{\mathrm{lin}}\left(  \mathbf{r}\right)  =$ $j_{0}\left(
\mathbf{r}\right)  $ . The relative coefficient $-1$ in (\ref{??}) is chosen
so as to make the field also satisfy the other Maxwell equation $\nabla
\times\mathbf{E}^{\mathrm{lin}}\left(  \mathbf{r}\right)  =0$. Consequently,
the component $\mathbf{e}_{\theta}\cdot\mathbf{E}^{\mathrm{lin}}\left(
\mathbf{r}\right)  $ of the field (\ref{4}) tangent to the sphere is
continuous at the border of the sphere $r=R$, while its perpendicular
component $\mathbf{r}\cdot\mathbf{E}^{\mathrm{lin}}\left(  \mathbf{r}\right)
$ is not.

By doing the same calculations as in the previous Subsection (see Appendix
3)\ we find the nonlinear correction to the electric field of the dipole taken
in the approximation $\mathbf{E}\left(  \mathbf{r}\right)  \simeq
\mathbf{E}^{\mathrm{lin}}\left(  \mathbf{r}\right)  $ (now $\theta$ is the
angle between $\mathbf{r}$ and $\mathbf{p)}:$%
\begin{align}
&  \frac{-2R^{9}}{\mathfrak{L}_{\mathfrak{FF}}p^{3}}\mathbf{E}^{\mathrm{nl}%
}\left(  \mathbf{r}\right)
=\label{Nonlinear electric field correction with full terms}\\
&  \left\{  \left[  \left(  \allowbreak-\frac{13}{15}+\frac{324}{385}%
\frac{r^{2}}{R^{2}}\right)  \cos\theta-\frac{108}{77}\frac{r^{2}}{R^{2}}%
\cos^{3}\theta\right]  \mathbf{e}_{r}\right. \nonumber\\
&  \left.  +\left[  \left(  \allowbreak\frac{13}{15}+\frac{432}{385}%
\frac{r^{2}}{R^{2}}\right)  \sin\theta-\frac{108}{77}\frac{r^{2}}{R^{2}}%
\sin^{3}\theta\right]  \mathbf{e}_{\theta}\right\}  \Theta\left(  R-r\right)
\nonumber\\
&  +\left\{  \left[  \left(  \allowbreak\frac{2}{5}\frac{R^{3}}{r^{3}}%
+\frac{18}{35}\frac{R^{5}}{r^{5}}\allowbreak\allowbreak+\frac{68}{33}%
\frac{R^{9}}{r^{9}}\right)  \cos\theta+\left(  -\frac{6}{7}\frac{R^{5}}{r^{5}%
}+\frac{60}{11}\frac{R^{9}}{r^{9}}\right)  \cos^{3}\theta\right]
\mathbf{e}_{r}\right. \nonumber\\
&  \left.  +\left[  \left(  \frac{1}{5}\frac{R^{3}}{r^{3}}-\frac{18}{35}%
\frac{R^{5}}{r^{5}}+\frac{76}{33}\frac{R^{9}}{r^{9}}\right)  \sin
\theta+\left(  \frac{9}{14}\frac{R^{5}}{r^{5}}\allowbreak-\allowbreak\frac
{45}{22}\frac{R^{9}}{r^{9}}\right)  \sin^{3}\theta\right]  \mathbf{e}_{\theta
}\right\}  \Theta\left(  r-R\right)  .\nonumber
\end{align}

At large distances$\mathfrak{\ }$we have%
\begin{equation}
\left.  \mathbf{E}\left(  \mathbf{r}\right)  \right\vert _{r>>R}%
=\mathbf{E}^{\text{lin}}\left(  \mathbf{r}\right)  +\left.  \mathbf{E}%
^{\mathrm{nl}}\left(  \mathbf{r}\right)  \right\vert _{r>>R}=\mathbf{E}%
^{\text{lin}}\left(  \mathbf{r}\right)  \left(  1-\frac{1}{10}\mathfrak{L}%
_{\mathfrak{FF}}\left(  \frac{p}{R^{3}}\right)  ^{2}\right)  ,
\label{Nonlinear correction for electric}%
\end{equation}
where $\mathbf{E}^{\text{lin}}\left(  \mathbf{r}\right)  =\mathbf{E}%
^{>}\left(  \mathbf{r}\right)  =\frac{p}{r^{3}}\left(  2\cos\theta
\mathbf{e}_{r}+\sin\theta\mathbf{e}_{\theta}\right)  $ is the (outside) linear
electric field (\ref{64}). With the Euler-Heisenberg Lagrangian we use
(\ref{LFF}) for $\mathfrak{L}_{\mathfrak{FF}}.$ Then the overall field in QED
is given by
\begin{equation}
\left.  \mathbf{E}\left(  \mathbf{r}\right)  \right\vert _{r>>R}%
=\mathbf{E}^{\text{lin}}\left(  \mathbf{r}\right)  \left(  1-\frac{2}{45}%
\frac{\alpha}{45\pi}\left(  \frac{E^{\mathrm{<}}}{E^{\mathrm{Sch}}}\right)
^{2}\right)  , \label{Economical notation}%
\end{equation}
where $E^{\mathrm{Sch}}$ is Schwinger's characteristic value for electric
field. The form (\ref{Economical notation}) is valid both in HL and Gauss
systems of units.

The statements, made about the magnetic moment $\boldsymbol{\mathcal{M}}$\ and
its magnetic field at the end of the previous Subsection, can be repeated as
applied to the electric moment $\mathbf{p}$ and its electric field. For
instance, the analog of (\ref{self-coupling}) is the equation for
selfinteracting electric dipole model%
\begin{equation}
\mathbf{p}^{\text{nlc}}=\mathbf{p}-\frac{\mathbf{p}^{\text{nlc}}}%
{10}\mathfrak{L}_{\mathfrak{FF}}\left(  \frac{p^{\text{nlc}}}{R^{3}}\right)
^{2}, \label{self-coupling2}%
\end{equation}
where $\mathbf{p}$ is the "bare" moment, introduced in (\ref{1}), and
$\mathbf{p}^{\text{nlc}}$ is the nonlinerly corrected moment.

The vast difference of the results (\ref{Nonlinear correction for electric})
-- (\ref{Economical notation}) of the present Subsection from those relating
to the spherical-symmetric case of Subsection 4.1.1, namely, from Eqs.
(\ref{Symmetric nonlinear correction}), (\ref{Euler-Heisenberg}), is that the
correction in (\ref{Nonlinear correction for electric}) --
(\ref{Economical notation}) contains only internal properties of the source,
and does not depend on the distance from it, in contrast to
(\ref{Symmetric nonlinear correction}), (\ref{Euler-Heisenberg}). Eqs.
(\ref{Nonlinear correction for electric}) -- (\ref{Economical notation}) may
be thought of as a sort of renormalization of the dipole moment $p$, whereas
(\ref{Euler-Heisenberg}) is not a correction to the charge $q,$ but rather the
field renormalization. The same remark is valid also for the magnetic dipole
of the previous Subsection, Eqs. (\ref{>>}) -- (\ref{Mnlc}).

\section{Some numerical estimates}

\bigskip The nonlinear corrections to electromagnetic fields in the blank
vacuum considered in the previous sections are cubic with respect to the
primary, linear fields. This means that these are most essential for large
fields. In this section we discuss numerically a few important instances of
strong fields.

\subsection{Baryons}

The of the neutron magnetic moment in nuclear magnetons $\mu_{N}=\frac
{e}{2m_{\text{p}}},$ where $m_{\text{p}}$ is the proton mass (remind that, in
Gauss units, $\frac{e\hbar}{2m_{\text{p}}c}=5.05078324\cdot10^{-24}$ $%
\operatorname{erg}%
$/$%
\operatorname{G}%
)$\ is measured \cite{Nakamura} with the high precision of $\sigma
_{\mathcal{M}}=0.0000005\mu_{N}$\ to be
\begin{equation}
\mathcal{M}=-1.9130427\mu_{N}%
\end{equation}

The magnetic radius of the neutron is believed to make (up to the third
decimal point)%
\begin{equation}
R=0.862\text{ f}%
\operatorname{m}%
. \label{R}%
\end{equation}

Linear magnetic field (\ref{magdip>}) of the neutron due to its magnetic
moment at a distance $r$\ (along the magnetic moment direction) is
$B^{\text{lin}}=2\frac{\mathcal{M}}{r^{3}}.$ It reaches Schwinger's critical
value, equal to $B_{\text{Sch}}=4.4\times10^{13}%
\operatorname{G}%
$ in Gauss units, at $r=r_{\text{c}}=7.55$ $%
\operatorname{fm}%
$.

Then, the field exceeds Schwinger's value, when we are closer to the neutron
than its $7.55/0.86=$ $8.8$ radii.

The magnetization of a neutron, defined within the simple model of Subsection
4.2.1 as $M=\frac{3\mathcal{M}}{4\pi R^{3}}$ makes in Gauss units%

\begin{equation}
M=3.75\times10^{15}%
\operatorname{G}%
=852B_{\text{Sch}}, \label{Neutron magnetization}%
\end{equation}
and the magnetic field (\ref{magdip<}) at its electromagnetic radius (\ref{R})
is about $3\cdot10^{16}%
\operatorname{G}%
.$ The value (\ref{Neutron magnetization}) is of the order of magnetization of
a strongest magnetar field and may be understood as magnetization of a neutron
star, provided the latter is viewed upon as a spontaneously magnetized dense
matter of paralelly arranged magnetic dipoles, neutrons, if they are as close
to each other as their magnetic radius (\ref{R}). (The corresponding density
of matter would be 6.2$\cdot10^{14}$ $%
\operatorname{g}%
/%
\operatorname{cm}%
^{3}).$ For the proton the value of magnetization is twice as large,
$7.5\times10^{16}$ $%
\operatorname{G}%
$ $=1.7\cdot10^{3}B_{\text{Sch}}.$

\bigskip With so large magnetization, the nonlinear correction in
(\ref{Mnlc}), calculated with the use of (\ref{LFF})%
\begin{equation}
\frac{7}{5}\mathfrak{L}_{\mathfrak{FF}}\left(  \frac{\mathcal{M}}{R^{3}%
}\right)  ^{2}=\frac{28}{225\pi}\alpha\left(  \frac{M}{B_{\text{Sch}}}\right)
^{2}=2.89\left(  \frac{1.9\cdot5.05}{4.4(0.86)^{3}}\right)  ^{2}=33.5>>1,
\label{33.5}%
\end{equation}
takes us far out of the scope of validity of the power expansion
(\ref{General current}). Therefore, for treating the neutron and proton, as
well as the most of other baryons, we must find a way of going beyond this
framework. While leaving this task for future work, we, however, can indicate
now that the results of Subsubsection 4.2.1 are ready for service when
particles or resonances with smaller magnetic moments are concerned. We
attract attention to the excited baryon $\Xi^{\ast0},$ whose magnetic moment
is estimated \cite{Lee} at the level of $0.16(4),$ and to the $\Delta^{0}$
decuplet member, with the magnetic moment $-0.035(2).$Then the nonlinear
correction to magnetization (and to its magnetic moment as well, assuming the
magnetic radius being of the same order as that of the neutron) makes
$0.23<<1$ for $\Xi^{\ast0}$ and $0.01$ $<<1$ for $\Delta^{0}.$ The correction
for $\Xi^{\ast0}$ is within the range of admitted theoretical errors in Ref.
\cite{Lee}, that reports the calculations of magnetic moments appealing to the
quark structure and fulfilled within the lattice QCD with the extrapolation to
the observable value of the pion mass (with the help of the chiral
perturbation theory). As for $\Delta^{0},$ the above nonlinear correction
$0.01$ somewhat falls off of the admitted range of $2/35=0.057.$ But other
calculations (those of Ref. \cite{Butler} made with the help of chiral
perturbation theory) give $-0.17(4)$ for the magnetic moment of \ $\Delta
^{0},$ and, correspondingly, produce the nonlinear correction of $0.26,$ while
the theoretical indeterminacy is $0.23.$ Another candidate treatable with our
formulas may be $\Sigma^{\ast0}.$ Experimental values do not exist for these
baryons, but there are theoretical predictions within the same theory that
produces good coincidence with experimentally measured magnetic moments
wherever the latter are available, neutron and proton included.

We conclude that the nonlinear correction to the magnetic moment of the above
baryons, as lying within the range of existing theoretical indeterminacy and,
perhaps, within its discrepancy with a future experiment, indicates that it
may come seriously into play already at the very next step of improving the
theoretical results. \bigskip

\subsection{Pulsars}

\bigskip

Near the surface at the north magnetic pole of a fast rotating radio-pulsar
\cite{Beskin} (also for certain magnetars \cite{Magnetar}) we may accept that
the magnetic field is close to Schwinger's value, $B^{<}\simeq B_{\text{Sch}%
}.$ Then at large distances from the surface the relative nonlinear correction
to the magnetic field (the second term in the brace of Eq. (\ref{Final field}%
)) is $\simeq7\cdot10^{-5},$ the absolute correction to the field being thus
$\simeq3\cdot10^{9}%
\operatorname{G}%
.$ Some magnetars \cite{Kouv} (soft gamma-ray repeaters, anomaleous X-ray
pulsars \textit{etc}.) are believed to have the magnetic fields up to
$10^{15}$G, (and, under certain assumptions, even two orders larger
\cite{Usov}). Then, in absolute value the nonlinear correction may achieve the
huge value of somewhat below $10^{14}$G, which may influence the fundamental
physical processes in the neutron star magnetosphere and hence the mechanisms
of their radiation.

We can rely on Eqs. (\ref{>>}) -- (\ref{Mnlc}) till the correction in them
does not exceed, say, 10\%. For this case Fig. 1, drawn following Eq.
(\ref{Nonlinear magnetic field correction with full terms}) and corresponding
to the field at the surface equal to $B^{<}=37B_{\text{Sch}},$ shows how the
ratio between the nonlinear field and the linear field behaves as the distance
to the surface increases. Whereas at large distances from the pulsar the
corrected field follows the same magnetic dipole law (\ref{magdip>}) as the
primary, linear field, near the surface it considerably deviates from the
field of magnetic dipole. At large distances, this ratio is constant, and the
nonlinear effect manifests itself as a renormalization of the observed dipole
field and its magnetic moment.

\bigskip

\bigskip

\bigskip

\bigskip

\bigskip

%

\begin{center}
\includegraphics[
trim=0.000000in 0.000000in 0.000000in 0.574914in,
height=1.7841in,
width=2.4431in
]%
{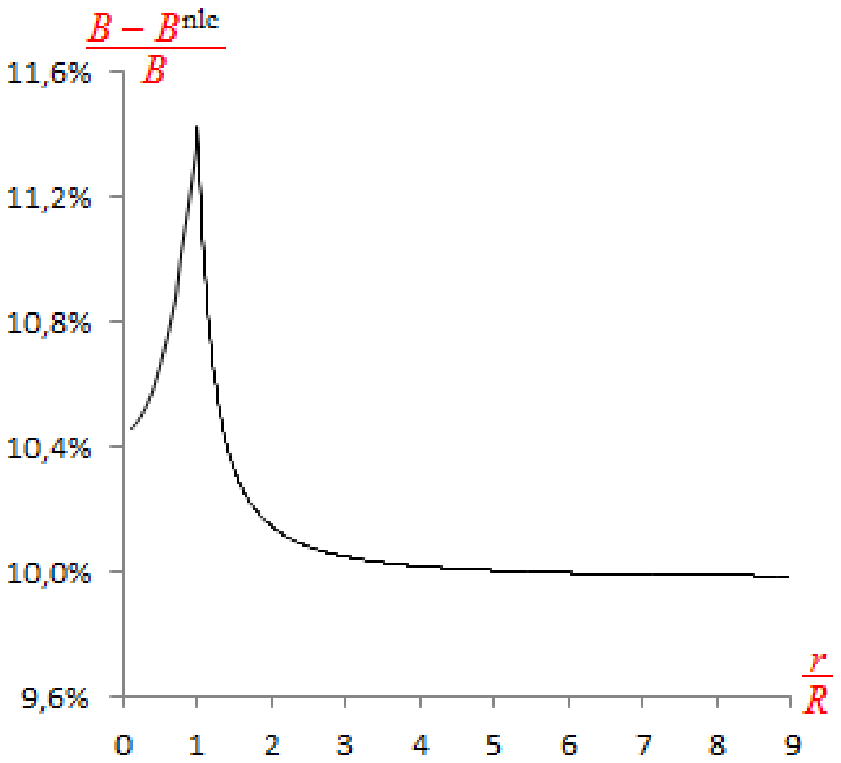}%
\\
FIG 1. Nonlinear magnetic corrections to the field for Magnetars
\end{center}

For larger magnetization of a magnetar, which might, within the concept of its
purely ferromagnetic neutron star nature, ultimately reach the magnetization
of the neutron (\ref{Neutron magnetization}), we fall beyond the applicability
of Eq. (\ref{>>}) -- (\ref{Mnlc}), as discussed in connection with
(\ref{33.5}) above.

\section{Conclusion}

In this paper we considered nonlinear Maxwell equations (\ref{General current}%
), (\ref{a}) -- (\ref{nonlincur}) in QED and other nonlinear Abelian gauge
theories without external field by including the nonlinearity to the third
power with respect to the field. The nonlinearity was taken into account
within the low-frequency, low-momentum approximation for the off-shell
four-prong photon diagram, to which end the 4-rank polarization tensor
(\ref{Fourth rank polarizator tensor}), (\ref{Constant tensor}) (also in
external field (\ref{Fourth rank tensor})) is calculated as the fourh
derivative over the 4-potential of the effective action taken in the local
approximation, i.e. the one, where its possible dependence upon the space-time
derivatives of the field strength is neglected. All the results are expressed
in terms of the second derivatives of the local effective action over the two
field invariants. In QED the role of this effective action is played in our
calculations by the one-loop Euler-Heisenberg action.

The nonlinearly-induced current (\ref{Non linear induced current}) to be
treated as a source in the standard Maxwell equations
(\ref{Maxwell's Equation}) is expressed in terms of the field strengths as
(\ref{CAIO ANSWER 2: FIX SIGN ON L_GG}). The field equations are written
separately for the special cases when a static electric
(\ref{Solution in terms of projection operator}), (\ref{Cubic electric field})
or static magnetic (\ref{Bnl}), (\ref{Cubic magnetic field}) field is present
alone. The projection operators $\frac{\nabla_{i}\nabla_{j}}{\nabla^{2}}$ and
$\delta_{ij}-\frac{\nabla_{i}\nabla_{j}}{\nabla^{2}}$ involved in these cubic
integro-differential equations become trivial, when higher symmetries are
prescribed to the solutions. Such are the cases of the electric field produced
by a sperically symmetric charge distribution, charged finite-thickness plane
and straight-linear thread, and the case of the magnetic field produced by a
straight current-carrying wire. The above cubic equations, when treated
perturbatively, give nonlinear corrections to the corresponding fields,
(\ref{Euler-Heisenberg}), (\ref{46a}), (\ref{46b}), (\ref{Euler-Heisenberg2}).
For a lower, cylindric symmetry of the field produced by the homogeneously
magnetized sphere (\ref{j}), and by homogeneously electrically polarized
sphere (\ref{1}), the no longer trivial projection operators are realized
using spherical harmonic expansion to lead to the following results: the
primary dipole electric and magnetic fields reproduce themselves at large
distances after the nonlinearity is switched on, the electric and magnetic
moments undergoing nonlinear renormalizations. Equations for self-coupling of
electric and magnetic dipole moments are (\ref{self-coupling2}) and
(\ref{self-coupling}), respectively. No spontaneous electrization, nor
magnetization occurs.

Numerical estimates in QED witness that the magnetic field due to the magnetic
moment of the neutron is to large at its surface to be treated in the
framework of the power expansion with respect to the field. On the contrary,
magnetic moments of the excited baryon $\Xi^{\ast0}$ and the $\Delta^{0}$
decuplet member do fit the cubic approximation, and the corresponding
nonlinear corrections to them lie within the limits admitted by the
theoretical and experimental indeterminacies.

In pulsars and magnetars the relative nonlinear corrections to their magnetic
moments do not exceed magnitudes admitted by the validity of the cubic
approximation as long as the surface magnetic field remains below
1.6$\cdot10^{15}%
\operatorname{G}%
$ (in Gauss units). Whereas the nonlinearity just renormalizes the value of
the surface magnetic field estimated by an Earth observer, it causes the
difference with the "unrenormalized" local value that may be obtained using
certain theoretical models about the origin of the magnetic field of a neutron
star. The absolute difference may achieve the huge value of up to
$\cdot10^{14}%
\operatorname{G}%
-$ not going beyond the approximation. Surely, many physical processes in a
magnetic field are sensitive to such changes.

\section*{Acknowledgements}

C. Costa acknowledges the support of CAPES. D. Gitman thanks FAPESP and CNPq
for permanent support and is grateful for the support from the Project
2.3684.2011 of Tomsk State University and FTP, contract No 14.B37.21.0911. A.
Shabad acknowledges the support of FAPESP, Processo 2011/51867-9, and of RFBR
under the Project 11-02-00685-a. He also thanks USP for kind hospitality
extended to him during his stay in Sao Paulo, Brazil, where this work was
partially fulfilled. The authors are thankful to T. Adorno for some important discussions.

\section*{Appendix 1}

Here we list the variational derivatives of the effective action $\Gamma$
(\ref{Gamma}) assuming that the Lagrangian $\mathfrak{L}$ does not depend upon
derivatives of the field strengths. All the equations below are understood as
reduced to a space- and time-independent background field $F_{\alpha\beta
}\longrightarrow\mathcal{F}_{\alpha\beta}.$ \ \ 

The second-order derivative is:%
\begin{align}
&  \frac{\delta^{2}\Gamma}{\delta A^{\mu}(x)\delta A^{\tau}(y)}=\int
\mathrm{d}^{4}z\left\{  \frac{\partial\mathfrak{L}}{\partial\mathfrak{F}%
}\left(  \eta_{\mu\tau}\eta_{\alpha\beta}-\eta_{\mu\beta}\eta_{\alpha\tau
}\right)  \right.  +\nonumber\\
&  +\frac{\partial\mathfrak{L}}{\partial\mathfrak{G}}\epsilon_{\alpha\mu
\beta\tau}+\frac{\partial^{2}\mathfrak{L}}{\partial\mathfrak{F}^{2}}%
F_{\alpha\mu}F_{\beta\tau}+\frac{\partial^{2}\mathfrak{L}}{\partial
\mathfrak{G}^{2}}\tilde{F}_{\alpha\mu}\tilde{F}_{\beta\tau}+\nonumber\\
&  +\left.  \frac{\partial^{2}\mathfrak{L}}{\partial\mathfrak{F}%
\partial\mathfrak{G}}\left[  {F}_{\alpha\mu}\tilde{F}_{\beta\tau}+\tilde
{F}_{\alpha\mu}{F}_{\beta\tau}\right]  \right\}  \left(  \frac{\partial
}{\partial z_{\alpha}}\delta^{4}(x-z)\right)  \left(  \frac{\partial}{\partial
z_{\beta}}\delta^{4}(y-z)\right)  .\quad\label{secdir}%
\end{align}

The third order-derivative is (note that the $F_{\gamma\sigma}$ tensor, which
gives zero in the no-background field case dealt with in the body of the
paper, appears in every term):%
\begin{align}
&  \frac{\delta^{3}\Gamma}{\delta A^{\mu}(x)\delta A^{\tau}(y)\delta
A^{\sigma}(u)}=\label{thirdir}\\
&  \int\mathrm{d}^{4}z\left\{  \left(  \eta_{\mu\tau}\eta_{\alpha\beta}%
-\eta_{\mu\beta}\eta_{\alpha\tau}\right)  \left[  \frac{\partial
^{2}\mathfrak{L}}{\partial\mathfrak{F}^{2}}F_{\gamma\sigma}+\frac{\partial
^{2}\mathfrak{L}}{\partial\mathfrak{G}\partial\mathfrak{F}}\tilde{F}%
_{\gamma\sigma}\right]  \frac{\partial}{\partial z_{\gamma}}\delta
^{4}(u-z)+\right. \nonumber\\
&  +\epsilon_{\alpha\mu\beta\tau}\left[  \frac{\partial^{2}\mathfrak{L}%
}{\partial\mathfrak{F}\partial\mathfrak{G}}F_{\gamma\sigma}+\frac{\partial
^{2}\mathfrak{L}}{\partial\mathfrak{G}^{2}}\tilde{F}_{\gamma\sigma}\right]
\frac{\partial}{\partial z_{\gamma}}\delta^{4}(u-z)\nonumber\\
&  +\left[  \frac{\partial^{3}\mathfrak{L}}{\partial\mathfrak{F}^{3}}%
F_{\alpha\mu}F_{\beta\tau}F_{\gamma\sigma}+\frac{\partial^{3}\mathfrak{L}%
}{\partial\mathfrak{G}\partial\mathfrak{F}^{2}}F_{\alpha\mu}F_{\beta\tau
}\tilde{F}_{\gamma\sigma}\right]  \frac{\partial}{\partial z_{\gamma}}%
\delta^{4}(u-z)+\nonumber\\
&  +\frac{\partial^{2}\mathfrak{L}}{\partial\mathfrak{F}^{2}}\left[
F_{\alpha\mu}\left(  \eta_{\tau\sigma}\frac{\partial}{\partial z^{\beta}}%
-\eta_{\beta\sigma}\frac{\partial}{\partial z^{\tau}}\right)  +F_{\beta\tau
}\left(  \eta_{\mu\sigma}\frac{\partial}{\partial z^{\alpha}}-\eta
_{\alpha\sigma}\frac{\partial}{\partial z^{\mu}}\right)  \right]  \delta
^{4}(u-z)\emph{+}\nonumber\\
&  \emph{+}\tilde{F}_{\alpha\mu}\tilde{F}_{\beta\tau}\left[  \frac
{\partial^{3}\mathfrak{L}}{\partial\mathfrak{F}\partial\mathfrak{G}^{2}%
}F_{\gamma\sigma}+\frac{\partial^{3}\mathfrak{L}}{\partial\mathfrak{G}^{3}%
}\tilde{F}_{\gamma\sigma}\right]  \frac{\partial}{\partial z_{\gamma}}%
\delta^{4}(u-z)\emph{+}\quad\nonumber\\
&  +\frac{\partial^{2}\mathfrak{L}}{\partial\mathfrak{G}^{2}}\left[  \tilde
{F}_{\alpha\mu}\epsilon_{\beta\tau\gamma\sigma}+\tilde{F}_{\beta\tau}%
\epsilon_{\alpha\mu\gamma\sigma}\right]  \frac{\partial}{\partial z_{\gamma}%
}\delta^{4}(u-z)+\nonumber\\
&  +\frac{\partial^{2}\mathfrak{L}}{\partial\mathfrak{F}\partial\mathfrak{G}%
}[{F}_{\alpha\mu}\epsilon_{\beta\tau\gamma\sigma}\frac{\partial}{\partial
z_{\gamma}}+\tilde{F}_{\beta\tau}\left(  \eta_{\mu\sigma}\frac{\partial
}{\partial z^{\alpha}}-\eta_{\alpha\sigma}\frac{\partial}{\partial z^{\mu}%
}\right)  +\tilde{F}_{\alpha\mu}\left(  \eta_{\tau\sigma}\frac{\partial
}{\partial z^{\beta}}-\eta_{\beta\sigma}\frac{\partial}{\partial z^{\tau}%
}\right)  +\nonumber\\
&  +F_{\beta\tau}\epsilon_{\alpha\mu\gamma\sigma}\frac{\partial}{\partial
z_{\gamma}}]\delta^{4}(u-z)+\nonumber\\
&  +\left[  {F}_{\alpha\mu}\tilde{F}_{\beta\tau}+\tilde{F}_{\alpha\mu}%
{F}_{\beta\tau}\right]  \left[  \frac{\partial^{3}\mathfrak{L}}{\partial
\mathfrak{F}^{2}\partial\mathfrak{G}}F_{\gamma\sigma}+\frac{\partial
^{3}\mathfrak{L}}{\partial\mathfrak{F}\partial\mathfrak{G}^{2}}\tilde
{F}_{\gamma\sigma}\right]  \text{ \ }\nonumber\\
&  \left.  \frac{\partial}{\partial z_{\gamma}}\delta^{4}(u-z)\right\}
\left(  \frac{\partial}{\partial z_{\alpha}}\delta^{4}(x-z)\right)  \left(
\frac{\partial}{\partial z_{\beta}}\delta^{4}(y-z)\right)  .\nonumber
\end{align}

The fourth-order polarization tensor, responsible for the first nonlinear
correction in the no-external-field case is:%
\begin{align}
&  \frac{\delta^{4}\Gamma}{\delta A^{\mu}\left(  x\right)  \delta A^{\tau
}\left(  y\right)  \delta A^{\sigma}\left(  u\right)  \delta A^{\rho}\left(
v\right)  }=\label{Fourth rank tensor}\\
&  =\int\mathrm{d}^{4}z\text{ }\left\{  \frac{\partial^{2}\mathfrak{L}%
}{\partial\mathfrak{F}^{2}}\left[  \left(  \eta_{\alpha\lambda}^{\text{ }}%
\eta_{\rho\mu}-\eta_{\mu\lambda}^{\text{ }}\eta_{\alpha\rho}\right)  \left(
\eta_{\tau\sigma}^{\text{ }}\eta_{\beta\gamma}-\eta_{\beta\sigma}^{\text{ }%
}\eta_{\tau\gamma}\right)  \right.  \right. \nonumber\\
&  \left.  +\left(  \eta_{\beta\lambda}^{\text{ }}\eta_{\rho\tau}-\eta
_{\tau\lambda}^{\text{ }}\eta_{\beta\rho}\right)  \left(  \eta_{\mu\sigma
}^{\text{ }}\eta_{\alpha\gamma}-\eta_{\alpha\sigma}^{\text{ }}\eta_{\mu\gamma
}\right)  +\left(  \eta_{\gamma\lambda}^{\text{ }}\eta_{\rho\sigma}%
-\eta_{\sigma\lambda}^{\text{ }}\eta_{\gamma\rho}\right)  \left(  \eta
_{\mu\tau}^{\text{ }}\eta_{\alpha\beta}-\eta_{\mu\beta}^{\text{ }}\eta
_{\alpha\tau}^{\text{ }}\right)  \right] \nonumber\\
&  +\frac{\partial^{2}\mathfrak{L}}{\partial\mathfrak{F}\partial\mathfrak{G}%
}\left[  \left(  \eta_{\mu\tau}^{\text{ }}\eta_{\alpha\beta}-\eta_{\mu\beta
}^{\text{ }}\eta_{\alpha\tau}^{\text{ }}\right)  \epsilon_{\lambda\rho
\gamma\sigma}+\left(  \eta_{\gamma\lambda}^{\text{ }}\eta_{\rho\sigma}%
-\eta_{\sigma\lambda}^{\text{ }}\eta_{\gamma\rho}\right)  \epsilon_{\alpha
\mu\beta\tau}+\left(  \eta_{\alpha\lambda}^{\text{ }}\eta_{\rho\mu}-\eta
_{\mu\lambda}^{\text{ }}\eta_{\alpha\rho}\right)  \epsilon_{\beta\tau
\gamma\sigma}\right. \nonumber\\
&  \left.  +\left(  \eta_{\beta\lambda}^{\text{ }}\eta_{\rho\tau}-\eta
_{\tau\lambda}^{\text{ }}\eta_{\beta\rho}\right)  \epsilon_{\alpha\mu
\gamma\sigma}+\left(  \eta_{\mu\sigma}^{\text{ }}\eta_{\alpha\gamma}%
-\eta_{\alpha\sigma}^{\text{ }}\eta_{\mu\sigma}^{\text{ }}\right)
\epsilon_{\lambda\rho\beta\tau}+\left(  \eta_{\tau\sigma}^{\text{ }}%
\eta_{\beta\gamma}-\eta_{\beta\sigma}^{\text{ }}\eta_{\tau\gamma}^{\text{ }%
}\right)  \epsilon_{\lambda\rho\alpha\mu}\right] \nonumber\\
&  +\frac{\partial^{2}\mathfrak{L}}{\partial\mathfrak{G}^{2}}\left[
\epsilon_{\alpha\mu\beta\tau}\epsilon_{\lambda\rho\gamma\sigma}+\epsilon
_{\lambda\rho\alpha\mu}\epsilon_{\beta\tau\gamma\sigma}+\epsilon_{\lambda
\rho\beta\tau}\epsilon_{\alpha\mu\gamma\sigma}\right] \nonumber\\
&  +\frac{\partial^{3}\mathfrak{L}}{\partial\mathfrak{F}^{3}}\left[
F_{\alpha\mu}F_{\beta\tau}\left(  \eta_{\gamma\lambda}^{\text{ }}\eta
_{\rho\sigma}-\eta_{\sigma\lambda}^{\text{ }}\eta_{\gamma\rho}\right)
+F_{\beta\tau}F_{\gamma\sigma}\left(  \eta_{\alpha\lambda}^{\text{ }}%
\eta_{\rho\mu}-\eta_{\mu\lambda}^{\text{ }}\eta_{\alpha\rho}\right)
+F_{\alpha\mu}F_{\gamma\sigma}\left(  \eta_{\beta\lambda}^{\text{ }}\eta
_{\rho\tau}-\eta_{\tau\lambda}^{\text{ }}\eta_{\beta\rho}\right)  \right.
\nonumber\\
&  \left.  +F_{\alpha\mu}F_{\lambda\rho}\left(  \eta_{\tau\sigma}^{\text{ }%
}\eta_{\beta\gamma}-\eta_{\beta\sigma}^{\text{ }}\eta_{\tau\gamma}\right)
+F_{\beta\tau}F_{\lambda\rho}\left(  \eta_{\mu\sigma}^{\text{ }}\eta
_{\alpha\gamma}-\eta_{\alpha\sigma}^{\text{ }}\eta_{\mu\gamma}\right)
+F_{\gamma\sigma}F_{\lambda\rho}\left(  \eta_{\mu\tau}^{\text{ }}\eta
_{\alpha\beta}-\eta_{\mu\beta}^{\text{ }}\eta_{\alpha\tau}^{\text{ }}\right)
\right] \nonumber\\
&  +\frac{\partial^{3}\mathfrak{L}}{\partial\mathfrak{F}^{2}\partial
\mathfrak{G}}\left[  \left(  F_{\alpha\mu}\tilde{F}_{\beta\tau}+\tilde
{F}_{\alpha\mu}F_{\beta\tau}\right)  \left(  \eta_{\gamma\lambda}^{\text{ }%
}\eta_{\rho\sigma}-\eta_{\sigma\lambda}^{\text{ }}\eta_{\gamma\rho}\right)
+\left(  F_{\beta\tau}\tilde{F}_{\gamma\sigma}+F_{\gamma\sigma}\tilde
{F}_{\beta\tau}\right)  \left(  \eta_{\alpha\lambda}^{\text{ }}\eta_{\rho\mu
}-\eta_{\mu\lambda}^{\text{ }}\eta_{\alpha\rho}\right)  \right. \nonumber\\
&  +\left(  F_{\gamma\sigma}\tilde{F}_{\lambda\rho}+\tilde{F}_{\gamma\sigma
}F_{\lambda\rho}\right)  \left(  \eta_{\mu\tau}^{\text{ }}\eta_{\alpha\beta
}-\eta_{\mu\beta}^{\text{ }}\eta_{\alpha\tau}^{\text{ }}\right)  +\left(
F_{\alpha\mu}\tilde{F}_{\gamma\sigma}+F_{\gamma\sigma}\tilde{F}_{\alpha\mu
}\right)  \left(  \eta_{\beta\lambda}^{\text{ }}\eta_{\rho\tau}-\eta
_{\tau\lambda}^{\text{ }}\eta_{\beta\rho}\right) \nonumber\\
&  +F_{\alpha\mu}\tilde{F}_{\lambda\rho}\left(  \eta_{\tau\sigma}^{\text{ }%
}\eta_{\beta\gamma}-\eta_{\beta\sigma}^{\text{ }}\eta_{\tau\gamma}\right)
+F_{\beta\tau}\tilde{F}_{\lambda\rho}\left(  \eta_{\mu\sigma}^{\text{ }}%
\eta_{\alpha\gamma}-\eta_{\alpha\sigma}^{\text{ }}\eta_{\mu\gamma}\right)
\nonumber\\
&  \left.  +F_{\alpha\mu}F_{\beta\tau}\epsilon_{\lambda\rho\gamma\sigma
}+F_{\beta\tau}F_{\gamma\sigma}\epsilon_{\lambda\rho\alpha\mu}+F_{\gamma
\sigma}F_{\lambda\rho}\epsilon_{\alpha\mu\beta\tau}+F_{\alpha\mu}%
\epsilon_{\lambda\rho\beta\tau}\right] \nonumber\\
&  +\frac{\partial^{3}\mathfrak{L}}{\partial\mathfrak{F}\partial
\mathfrak{G}^{2}}\left[  \tilde{F}_{\alpha\mu}\tilde{F}_{\beta\tau}\left(
\eta_{\gamma\lambda}^{\text{ }}\eta_{\rho\sigma}-\eta_{\sigma\lambda}^{\text{
}}\eta_{\gamma\rho}\right)  +\tilde{F}_{\alpha\mu}\tilde{F}_{\lambda\rho
}\left(  \eta_{\tau\sigma}^{\text{ }}\eta_{\beta\gamma}-\eta_{\beta\sigma
}^{\text{ }}\eta_{\tau\gamma}^{\text{ }}\right)  \right. \nonumber\\
&  +\tilde{F}_{\beta\tau}\tilde{F}_{\lambda\rho}\left(  \eta_{\mu\sigma
}^{\text{ }}\eta_{\alpha\gamma}-\eta_{\alpha\sigma}^{\text{ }}\eta_{\mu\sigma
}^{\text{ }}\right)  +\tilde{F}_{\gamma\sigma}\tilde{F}_{\lambda\rho}\left(
\eta_{\mu\tau}^{\text{ }}\eta_{\alpha\beta}-\eta_{\mu\beta}^{\text{ }}%
\eta_{\alpha\tau}^{\text{ }}\right)  +\left(  F_{\gamma\sigma}\tilde
{F}_{\lambda\rho}+\tilde{F}_{\gamma\sigma}F_{\lambda\rho}\right)
\epsilon_{\alpha\mu\beta\tau}\nonumber\\
&  \left.  +F_{\gamma\sigma}\tilde{F}_{\alpha\mu}\epsilon_{\lambda\rho
\beta\tau}+F_{\gamma\sigma}\tilde{F}_{\beta\tau}\epsilon_{\lambda\rho\alpha
\mu}+F_{\alpha\mu}\tilde{F}_{\lambda\rho}\epsilon_{\beta\tau\gamma\sigma
}+F_{\beta\tau}\tilde{F}_{\lambda\rho}\epsilon_{\alpha\mu\gamma\sigma}\right]
+\frac{\partial^{4}\mathfrak{L}}{\partial\mathfrak{F}^{4}}\left[  F_{\alpha
\mu}F_{\beta\tau}F_{\gamma\sigma}F_{\lambda\rho}\right] \nonumber\\
&  +\frac{\partial^{4}\mathfrak{L}}{\partial\mathfrak{F}^{3}\partial
\mathfrak{G}}\left[  F_{\alpha\mu}F_{\beta\tau}F_{\gamma\sigma}\tilde
{F}_{\lambda\rho}+F_{\alpha\mu}F_{\beta\tau}F_{\lambda\rho}\tilde{F}%
_{\gamma\sigma}+F_{\alpha\mu}F_{\gamma\sigma}F_{\lambda\rho}\tilde{F}%
_{\beta\tau}+F_{\beta\tau}F_{\gamma\sigma}F_{\lambda\rho}\tilde{F}_{\alpha\mu
}\right] \nonumber\\
&  +\frac{\partial^{4}\mathfrak{L}}{\partial\mathfrak{F}^{2}\partial
\mathfrak{G}^{2}}\left[  F_{\gamma\sigma}F_{\lambda\rho}\tilde{F}_{\alpha\mu
}\tilde{F}_{\beta\tau}+F_{\alpha\mu}F_{\beta\tau}\tilde{F}_{\gamma\sigma
}\tilde{F}_{\lambda\rho}+F_{\alpha\mu}F_{\gamma\sigma}\tilde{F}_{\beta\tau
}\tilde{F}_{\lambda\rho}\right. \nonumber\\
&  \left.  +F_{\alpha\mu}F_{\lambda\rho}\tilde{F}_{\gamma\sigma}\tilde
{F}_{\beta\tau}+F_{\beta\tau}F_{\lambda\rho}\tilde{F}_{\gamma\sigma}\tilde
{F}_{\alpha\mu}+F_{\gamma\sigma}F_{\beta\tau}\tilde{F}_{\alpha\mu}\tilde
{F}_{\lambda\rho}\right] \nonumber\\
&  +\frac{\partial^{4}\mathfrak{L}}{\partial\mathfrak{F}\partial
\mathfrak{G}^{3}}\left[  F_{\alpha\mu}\tilde{F}_{\beta\tau}\tilde{F}%
_{\gamma\sigma}\tilde{F}_{\lambda\rho}+F_{\beta\tau}\tilde{F}_{\alpha\mu
}\tilde{F}_{\gamma\sigma}\tilde{F}_{\lambda\rho}+F_{\gamma\sigma}\tilde
{F}_{\alpha\mu}\tilde{F}_{\beta\tau}\tilde{F}_{\lambda\rho}+F_{\lambda\rho
}\tilde{F}_{\alpha\mu}\tilde{F}_{\beta\tau}\tilde{F}_{\gamma\sigma}\right]
\nonumber\\
&  \left.  +\frac{\partial^{4}\mathfrak{L}}{\partial\mathfrak{G}^{4}}\tilde
{F}_{\alpha\mu}\tilde{F}_{\beta\tau}\tilde{F}_{\gamma\sigma}\tilde{F}%
_{\lambda\rho}\right\}  \frac{\partial\delta^{4}\left(  x-z\right)  }{\partial
z_{\alpha}}\frac{\partial\delta^{4}\left(  y-z\right)  }{\partial z_{\beta}%
}\frac{\partial\delta^{4}\left(  u-z\right)  }{\partial z_{\gamma}}%
\frac{\partial\delta^{4}\left(  v-z\right)  }{\partial z_{\lambda}}.\nonumber
\end{align}

\bigskip

\section*{Appendix 2}

Here we continue the calculation of nonlinear correction to the magnetic
dipole field starting with Eq. (\ref{cubic}). Write the divergence of
(\ref{cubic}) in terms of the spherical harmonics $Y_{1}^{0}\left(
\Omega\right)  =\frac{1}{2}\sqrt{\frac{3}{\pi}}\cos\theta$ and $Y_{3}%
^{0}\left(  \Omega\right)  =\frac{1}{4}\sqrt{\frac{7}{\pi}}\left(  5\cos
^{3}\theta-3\cos\theta\right)  \ $as%
\begin{align}
&  \frac{-108}{\mathfrak{L}_{\mathfrak{FF}}M^{3}}\boldsymbol{\nabla}%
\cdot\boldsymbol{\mathfrak{h}}\left(  \mathbf{r}\right)  =-16\delta\left(
r-R\right)  \sqrt{\frac{\pi}{3}}Y_{1}^{0}\left(  \Omega\right)
\label{Divergence}\\
&  +\frac{R^{9}}{r^{9}}\delta\left(  r-R\right)  \left(  \frac{56}{5}%
\sqrt{\frac{\pi}{3}}Y_{1}^{0}\left(  \Omega\right)  +\frac{24}{5}\sqrt
{\frac{\pi}{7}}Y_{3}^{0}\left(  \Omega\right)  \right) \nonumber\\
&  +\frac{R^{9}}{r^{10}}\Theta\left(  r-R\right)  \left(  -72\sqrt{\frac{\pi
}{3}}Y_{1}^{0}\left(  \Omega\right)  -24\sqrt{\frac{\pi}{7}}Y_{3}^{0}\left(
\Omega\right)  \right)  .\nonumber
\end{align}

We separate variables to radial and angle (normalized spherical harmonics)
parts%
\begin{equation}
\frac{-108}{\mathfrak{L}_{\mathfrak{FF}}M^{3}}\nabla\cdot
\boldsymbol{\mathfrak{h}}\left(  \mathbf{r}\right)  =\sum_{j}R_{j}\left(
r\right)  Y_{j}^{0}\left(  \Omega\right)  , \label{Separation of variables}%
\end{equation}
where, using the orthogonality relation $\int Y_{j}^{0}\left(  \Omega\right)
Y_{l}^{\ast m}\left(  \Omega\right)  d\Omega=\delta_{jl}\delta_{0m}$ we get%
\begin{equation}
R_{j}\left(  r\right)  =\frac{-108}{\mathfrak{L}_{\mathfrak{FF}}M^{3}}%
\int\nabla\cdot\boldsymbol{\mathfrak{h}}\left(  \mathbf{r}\right)  Y_{j}%
^{0}\left(  \Omega\right)  \mathrm{d}\Omega. \label{Radius part}%
\end{equation}

With the use of (\ref{Divergence}) we obtain%
\begin{align}
&  R_{j}\left(  r\right)  =\left[  \left(  {\tiny -}16+\frac{56}{5}\frac
{R^{9}}{r^{9}}\right)  \delta\left(  r{\tiny -}R\right)  {\tiny -}%
72\frac{R^{9}}{r^{10}}\Theta\left(  r{\tiny -}R\right)  \right]  \sqrt
{\frac{\pi}{3}}\delta_{1j}\label{Explicit radius part}\\
&  +\left[  \frac{24}{5}\frac{R^{9}}{r^{9}}\delta\left(  r{\tiny -}R\right)
-24\frac{R^{9}}{r^{10}}\Theta\left(  r{\tiny -}R\right)  \right]  \sqrt
{\frac{\pi}{7}}\delta_{3j}.\nonumber
\end{align}

We write the Green function in the following way \cite{Arfken}:\
\begin{equation}
\frac{1}{\left\vert \mathbf{r-r}^{\prime}\right\vert }=%
{\displaystyle\sum\limits_{l=0}^{\infty}}
{\displaystyle\sum\limits_{m=-l}^{l}}
\frac{4\pi}{2l+1}Y_{l}^{m}\left(  \Omega\right)  Y_{l}^{\ast m}\left(
\Omega^{\prime}\right)  \left\{
\begin{array}
[c]{c}%
\frac{r^{\prime l}}{r^{l+1}},\text{ \ if }r>r^{\prime}\\
\frac{r^{l}}{r^{\prime l+1}},\text{ \ if }r<r^{\prime}%
\end{array}
\right.  . \label{Green function}%
\end{equation}

Let us calculate (\ref{Vector form for magnetic field}), using
(\ref{Cubic magnetic field}), (\ref{Separation of variables}),
(\ref{Explicit radius part}) and (\ref{Green function}):

One can separate the integrals in the two regions of the space%
\begin{align*}
&  \frac{-2\left(  4\pi\right)  ^{3}R^{9}}{\mathfrak{L}_{\mathfrak{FF}%
}\mathcal{M}^{3}}\mathbf{B}^{\mathrm{nl}}\left(  \mathbf{r}\right)
=\frac{-2\left(  4\pi\right)  ^{3}R^{9}}{\mathfrak{L}_{\mathfrak{FF}%
}\mathcal{M}^{3}}\boldsymbol{\mathfrak{h}}\left(  \mathbf{r}\right) \\
&  +\frac{1}{4\pi}\boldsymbol{\nabla}\int\left[  \int_{0}^{r}\sum
\limits_{j}R_{j}\left(  r^{\prime}\right)  Y_{j}^{0}\left(  \Omega^{\prime
}\right)  \left(
{\displaystyle\sum\limits_{l=0}^{\infty}}
{\displaystyle\sum\limits_{m=-l}^{l}}
\frac{4\pi}{2l+1}Y_{l}^{m}\left(  \Omega\right)  Y_{l}^{\ast m}\left(
\Omega^{\prime}\right)  \frac{r^{\prime l}}{\text{ \ }r^{l+1}}\right)
r^{\prime2}\mathrm{d}r^{\prime}\right.  .\\
&  \left.  +\int_{r}^{\infty}\sum\limits_{j}R_{j}\left(  r^{\prime}\right)
Y_{j}^{0}\left(  \Omega^{\prime}\right)  \left(
{\displaystyle\sum\limits_{l=0}^{\infty}}
{\displaystyle\sum\limits_{m=-l}^{l}}
\frac{4\pi}{2l+1}Y_{l}^{m}\left(  \Omega\right)  Y_{l}^{\ast m}\left(
\Omega^{\prime}\right)  \frac{r^{l}}{\text{ \ }r^{\prime l+1}}\right)
r^{\prime2}\mathrm{d}r^{\prime}\right]  \mathrm{d}\Omega^{\prime}.
\end{align*}

Therefore, integrating over the solid angle we obtain%
\begin{align*}
&  \frac{-2\left(  4\pi\right)  ^{3}R^{9}}{\mathfrak{L}_{\mathfrak{FF}%
}\mathcal{M}^{3}}\mathbf{B}^{\mathrm{nl}}\left(  \mathbf{r}\right)
=\frac{-2\left(  4\pi\right)  ^{3}R^{9}}{\mathfrak{L}_{\mathfrak{FF}%
}\mathcal{M}^{3}}\boldsymbol{\mathfrak{h}}\left(  \mathbf{r}\right) \\
&  +\sum\limits_{j}\frac{1}{2j+1}\boldsymbol{\nabla}\left(  \int_{0}^{r}%
R_{j}\left(  r^{\prime}\right)  \frac{r^{\prime j+2}}{r^{j+1}}\mathrm{d}%
r^{\prime}+\int_{r}^{\infty}R_{j}\left(  r^{\prime}\right)  \frac{r^{j}%
}{r^{\prime j-1}}\mathrm{d}r^{\prime}\right)  Y_{j}^{0}\left(  \Omega\right)
.
\end{align*}

Now we show the gradient in spherical coordinates%
\begin{align*}
&  \frac{-2\left(  4\pi\right)  ^{3}R^{9}}{\mathfrak{L}_{\mathfrak{FF}%
}\mathcal{M}^{3}}\mathbf{B}^{\mathrm{nl}}\left(  \mathbf{r}\right)
=\frac{-2\left(  4\pi\right)  ^{3}R^{9}}{\mathfrak{L}_{\mathfrak{FF}%
}\mathcal{M}^{3}}\boldsymbol{\mathfrak{h}}\left(  \mathbf{r}\right)  +\\
&  +\sum\limits_{j}\frac{1}{2j+1}\left\{  \frac{\partial}{\partial r}\left[
\int_{0}^{r}R_{j}\left(  r^{\prime}\right)  \frac{r^{\prime j+2}}{r^{j+1}%
}\mathrm{d}r^{\prime}+\int_{r}^{\infty}R_{j}\left(  r^{\prime}\right)
\frac{r^{j}}{r^{\prime j-1}}\mathrm{d}r^{\prime}Y_{j}^{0}\left(
\Omega\right)  \right]  \mathbf{e}_{r}\right. \\
&  \left.  +\frac{1}{r}\left[  \int_{0}^{r}R_{j}\left(  r^{\prime}\right)
\frac{r^{\prime j+2}}{r^{j+1}}\mathrm{d}r^{\prime}+\int_{r}^{\infty}%
R_{j}\left(  r^{\prime}\right)  \frac{r^{j}}{r^{\prime j-1}}\mathrm{d}%
r^{\prime}\frac{\partial}{\partial\theta}Y_{j}^{0}\left(  \Omega\right)
\right]  \mathbf{e}_{\theta}\right\}  .
\end{align*}

Now we act by the derivatives%
\begin{align}
&  \frac{-2\left(  4\pi\right)  ^{3}R^{9}}{\mathfrak{L}_{\mathfrak{FF}%
}\mathcal{M}^{3}}\mathbf{B}^{\mathrm{nl}}\left(  \mathbf{r}\right)
=\frac{-2\left(  4\pi\right)  ^{3}R^{9}}{\mathfrak{L}_{\mathfrak{FF}%
}\mathcal{M}^{3}}\boldsymbol{\mathfrak{h}}\left(  \mathbf{r}\right)
\label{Radial and angular}\\
&  +\sum\limits_{j}\frac{1}{2j+1}\left[  \Sigma_{j}\left(  r\right)  Y_{j}%
^{0}\left(  \Omega\right)  \mathbf{e}_{r}+\Xi_{j}\left(  r\right)
\frac{\partial}{\partial\theta}Y_{j}^{0}\left(  \Omega\right)  \mathbf{e}%
_{\theta}\right]  .\nonumber
\end{align}
where%
\begin{align}
&  \Sigma_{j}\left(  r\right)  =-\left(  j+1\right)  \int_{0}^{r}R_{j}\left(
r^{\prime}\right)  \frac{r^{\prime j+2}}{r^{j+2}}\mathrm{d}r^{\prime}%
+j\int_{r}^{\infty}R_{j}\left(  r^{\prime}\right)  \frac{r^{j-1}}{r^{\prime
j-1}}\mathrm{d}r^{\prime},\label{Radial and angular functions}\\
&  \text{and}\nonumber\\
&  \Xi_{j}\left(  r\right)  =\int_{0}^{r}R_{j}\left(  r^{\prime}\right)
\frac{r^{\prime j+2}}{r^{j+2}}\mathrm{d}r^{\prime}+\int_{r}^{\infty}%
R_{j}\left(  r^{\prime}\right)  \frac{r^{j-1}}{r^{\prime j-1}}\mathrm{d}%
r^{\prime}.\nonumber
\end{align}

Therefore, by (\ref{Explicit radius part})
\begin{align}
&  \Sigma_{j}\left(  r\right)  =\sqrt{\frac{\pi}{3}}\left[  -\frac{64}%
{5}\delta_{1j}+\frac{432}{55}\frac{r^{2}}{R^{2}}\delta_{3j}\right]
\Theta\left(  R-r\right) \nonumber\\
&  +\sqrt{\frac{\pi}{3}}\left[  \left(  \frac{168}{5}\frac{R^{3}}{r^{3}%
}-32\frac{R^{9}}{r^{9}}\right)  \delta_{1j}+\left(  \frac{24}{5}\frac{R^{5}%
}{r^{5}}-\frac{336}{11}\frac{R^{9}}{r^{9}}\right)  \delta_{3j}\right]
\Theta\left(  r-R\right)  ,\label{Solution for Sigma and Csi}\\
&  \text{and}\nonumber\\
&  \Xi_{j}\left(  r\right)  =\sqrt{\frac{\pi}{7}}\left[  -\frac{64}{5}%
\delta_{1j}+\frac{144}{55}\frac{r^{2}}{R^{2}}\delta_{3j}\right]  \Theta\left(
R-r\right) \nonumber\\
&  +\sqrt{\frac{\pi}{7}}\left[  \left(  -\frac{84}{5}\frac{R^{3}}{r^{3}%
}+4\frac{R^{9}}{r^{9}}\right)  \delta_{1j}+\left(  -\frac{6}{5}\frac{R^{5}%
}{r^{5}}+\allowbreak\frac{42}{11}\frac{R^{9}}{r^{9}}\right)  \delta
_{3j}\right]  \Theta\left(  r-R\right)  . \label{KSI}%
\end{align}

Substituting (\ref{cubic}) and (\ref{Solution for Sigma and Csi}), (\ref{KSI})
in (\ref{Radial and angular}) we obtain the nonlinear magnetic field
(\ref{Nonlinear magnetic field correction with full terms}).

\section*{Appendix 3}

Here we continue the calculation of nonlinear correction to the electric
dipole field starting with Eq. (\ref{4}).

The electric field of an electric dipole can be dealt with analogously to the
case of magnetic dipole, although we use the longitudinal projection instead
the transverse projection. The electric field (\ref{64}) can be also written
as%
\begin{equation}
\mathbf{E}^{\text{lin}}\left(  \mathbf{r}\right)  =-\frac{p}{4\pi R^{3}}%
\Theta\left(  R-r\right)  \mathbf{e}_{3}+\left(  \frac{3p}{4\pi r^{3}}%
\cos\theta\mathbf{e}_{r}-\frac{p}{4\pi r^{3}}\mathbf{e}_{3}\right)
\Theta\left(  r-R\right)  , \label{Electric dipole}%
\end{equation}

Let us calculate (\ref{Cubic electric field}) substituting $\mathbf{E}%
^{\text{lin}}\left(  \mathbf{r}\right)  $ for $\mathbf{E}\left(
\mathbf{r}\right)  $ .

\bigskip The product $\frac{\left(  4\pi R^{3}\right)  ^{3}}{p^{3}}%
\mathbf{E}\left(  \mathbf{r}\right)  E^{2}\left(  \mathbf{r}\right)  $ is
given by%
\begin{align}
&  -\frac{2\left(  4\pi\right)  ^{3}R^{9}}{\mathfrak{L}_{\mathfrak{FF}}p^{3}%
}\boldsymbol{\mathcal{E}}\left(  \mathbf{r}\right)  =\left(  -\cos
\theta\mathbf{e}_{r}+\sin\theta\mathbf{e}_{\theta}\right)  \Theta\left(
R-r\right) \label{Cubic'}\\
&  +\left[  \left(  2\frac{R^{9}}{r^{9}}\cos\theta+6\frac{R^{9}}{r^{9}}%
\cos^{3}\theta\right)  \mathbf{e}_{r}+\left(  4\frac{R^{9}}{r^{9}}\sin
\theta-3\frac{R^{9}}{r^{9}}\sin^{3}\theta\right)  \mathbf{e}_{\theta}\right]
\Theta\left(  r-R\right)  .\nonumber
\end{align}

By (\ref{cubic}), one can compare $\boldsymbol{\mathcal{E}}\left(
\mathbf{r}\right)  $ to $\boldsymbol{\mathfrak{h}}\left(  \mathbf{r}\right)  $%
\begin{equation}
\frac{2\left(  4\pi\right)  ^{3}R^{9}}{\mathfrak{L}_{\mathfrak{FF}}p^{3}%
}\boldsymbol{\mathcal{E}}\left(  \mathbf{r}\right)  =\frac{2\left(
4\pi\right)  ^{3}R^{9}}{\mathfrak{L}_{\mathfrak{FF}}\mathcal{M}^{3}%
}\boldsymbol{\mathfrak{h}}\left(  \mathbf{r}\right)  -9\mathbf{e}_{3}%
\Theta\left(  R-r\right)  . \label{P}%
\end{equation}

The electric field is calculated from
(\ref{Solution in terms of projection operator}) and (\ref{P})
\begin{align}
&  -\frac{2\left(  4\pi\right)  ^{3}R^{9}}{\mathfrak{L}_{\mathfrak{FF}}p^{3}%
}\mathbf{E}^{\mathrm{nl}}\left(  \mathbf{r}\right)  =-\frac{1}{4\pi
}\boldsymbol{\nabla}\int\frac{\boldsymbol{\nabla}^{\prime}\cdot\left[
\frac{2\left(  4\pi\right)  ^{3}R^{9}}{\mathfrak{L}_{\mathfrak{FF}}p^{3}%
}\boldsymbol{\mathcal{E}}\left(  \mathbf{r}^{\prime}\right)  \right]
}{\left\vert \mathbf{r-r}^{\prime}\right\vert }\mathrm{d}\mathbf{r}^{\prime
}\label{E and H}\\
&  =-\frac{2\left(  4\pi\right)  ^{3}R^{9}}{\mathfrak{L}_{\mathfrak{FF}%
}\mathcal{M}^{3}}\left(  \mathbf{B}^{\mathrm{nl}}\left(  \mathbf{r}\right)
-\boldsymbol{\mathfrak{h}}\left(  \mathbf{r}\right)  \right)  +\frac{1}{4\pi
}\boldsymbol{\nabla}\int\frac{\boldsymbol{\nabla}^{\prime}\cdot\left[
9\mathbf{e}_{3}\Theta\left(  R-r^{\prime}\right)  \right]  }{\left\vert
\mathbf{r-r}^{\prime}\right\vert }\mathrm{d}\mathbf{r}^{\prime}.\nonumber
\end{align}

It is easy to show that%
\begin{equation}
\boldsymbol{\nabla}\cdot\left[  9\mathbf{e}_{3}\Theta\left(  R-r\right)
\right]  =-9\delta\left(  r-R\right)  \cos\theta. \label{Diverg}%
\end{equation}

The second term of (\ref{Diverg}) can be calculated in the same way as in the
magnetic case, i.e., by expanding of the Green function in spherical
harmonics, and calculating the solid angle integral with the help of
orthogonality relations%
\begin{align*}
&  -9\frac{1}{4\pi}\boldsymbol{\nabla}\int\frac{\delta\left(  R-r^{\prime
}\right)  \cos\theta}{\left\vert \mathbf{r-r}^{\prime}\right\vert }%
\mathrm{d}\mathbf{r}^{\prime}\\
&  =-9\frac{1}{4\pi}\boldsymbol{\nabla}\int\left[  \int_{0}^{r}\frac
{\delta\left(  R-r^{\prime}\right)  }{\left\vert \mathbf{r-r}^{\prime
}\right\vert }r^{\prime2}\mathrm{d}r^{\prime}+\int_{r}^{\infty}\frac
{\delta\left(  R-r^{\prime}\right)  }{\left\vert \mathbf{r-r}^{\prime
}\right\vert }r^{\prime2}\mathrm{d}r^{\prime}\right]  \frac{1}{2}\sqrt
{\frac{\pi}{3}}Y_{1}^{0}\left(  \Omega^{\prime}\right)  \mathrm{d}%
\Omega^{\prime}\\
&  =-3\boldsymbol{\nabla}\left[  \left(  \int_{0}^{r}\delta\left(
R-r^{\prime}\right)  \frac{r^{\prime3}}{r^{2}}\mathrm{d}r^{\prime}+\int
_{r}^{\infty}\delta\left(  R-r^{\prime}\right)  r\mathrm{d}r^{\prime}\right)
\frac{1}{2}\sqrt{\frac{\pi}{3}}Y_{1}^{0}\left(  \Omega\right)  \right] \\
&  =-3\boldsymbol{\nabla}\left[  \left(  \frac{R^{3}}{r^{2}}\Theta\left(
r-R\right)  +r\Theta\left(  R-r\right)  \right)  \cos\theta\right] \\
&  =\left(  6\frac{R^{3}}{r^{3}}\Theta\left(  r-R\right)  -3\Theta\left(
R-r\right)  \right)  \cos\theta\mathbf{e}_{r}+\left(  3\frac{R^{3}}{r^{3}%
}\Theta\left(  r-R\right)  +3\Theta\left(  R-r\right)  \right)  \sin
\theta\mathbf{e}_{\theta}.
\end{align*}

Therefore,
\begin{align*}
&  -\frac{2\left(  4\pi\right)  ^{3}R^{9}}{\mathfrak{L}_{\mathfrak{FF}}p^{3}%
}\mathbf{E}^{\mathrm{nl}}\left(  \mathbf{r}\right)  =-\frac{2\left(
4\pi\right)  ^{3}R^{9}}{\mathfrak{L}_{\mathfrak{FF}}\mathcal{M}^{3}}\left(
\mathbf{B}^{\mathrm{nl}}\left(  \mathbf{r}\right)  -\boldsymbol{\mathfrak{h}%
}\left(  \mathbf{r}\right)  \right)  +\left(  6\frac{R^{3}}{r^{3}}%
\Theta\left(  r-R\right)  -3\Theta\left(  R-r\right)  \right)  \cos
\theta\mathbf{e}_{r}\\
&  +\left(  3\frac{R^{3}}{r^{3}}\Theta\left(  r-R\right)  +3\Theta\left(
R-r\right)  \right)  \sin\theta\mathbf{e}_{\theta}.
\end{align*}

Thus the expression for electric field
(\ref{Nonlinear electric field correction with full terms}) is obtained.


\begin{thebibliography}{99}                                                                                               %


\bibitem {lettNuovCim}A.E. Shabad, Lett. Nuov. Cim. \textbf{2}, 457 (1972),
Ann. Phys. (N.Y.) \textbf{90}, 166 (1975)

\bibitem {nature}A.E. Shabad and V.V. Usov, Nature \textbf{295}, 215 (1982)

\bibitem {Ass}A.E. Shabad and V.V. Usov, Astrophys. and Space Sci.
\textbf{117}, 309 (1985); \textit{ibid}\textbf{ 128}, 377 (1986); H. Herold,
H. Ruder and G. Wunner, Phys. Rev. Lett.\textbf{ 54}, 1452 (1985); D.J.
Thompson \textit{et al., }ApJ\textit{. }\textbf{436}, 229 (1994); V.V. Usov
and D.B. Melrose, ApJ, \textbf{464}, 306 (1996)

\bibitem {Harding}A.K. Harding and D. Lai, \emph{Physics of strongly
magnetized neutron stars, }Rep. Prog. Phys. \textbf{69}, 2631 (2006)

\bibitem {Vachaspati}T.Vachaspati, Phys. Lett. B\textbf{265}, 258(1991);
K.Enquist and P.Olesen, Phys. Lett. \textbf{B 319}, 178 (1993); D.Grasso and
H.R.Rubinstein, Phys. Rept. \textbf{348}, 163 (2001)

\bibitem {pulsars}R.C.Duncan and C.Thompson, Astro. Phys. J. \textbf{392}, L9
(1992); J.M.Lattimer and M.Prakash, Phys. Rept. \textbf{442}, 109 (2007)

\bibitem {Co and Usov}R.P. Negreiros, F. Weber, M. Malheiro, V. Usov, Phys.
Rev. D \textbf{80,} 083006 (2009); R.P. Negreiros, I.N.

Mishustin, S. Schramm, F. Weber, Phys. Rev. D \textbf{82}, 103010 (2010)

\bibitem {Kharzeev}D.E.Kharzeev, L.D.McLerran and H.J.Warringa, Nucl. Phys.
A\textbf{803}, 227 (2008); V.S.Skokov, A.Illarionov and V.Toneev, Int. J. Mod.
Phys. A\textbf{24}, 5925 (2009)

\bibitem {Simonov}Yu.A. Simonov, B.O. Kerbikov, M.A. Andreichikov,
\emph{Asymptotic freedom in strong magnetic field,} arXiv:1210.0227[hep-ph];
V.D. Orlovsky, V.I. Shevchenko, \emph{Towards a quantum theory of the chiral
magnetic effect, }Phys. Rev. D, \textbf{82}, 094032 (2010); M. A.
Andreichikov, V. D. Orlovsky, and Yu. A. Simonov,\ \emph{Quark-Antiquark
System in Ultra-Intense Magnetic Field, }arXiv:1211.6568 [hep-ph]; Yu.A.
Simonov,\emph{ Neutral 3-body system in a strong magnetic field: Factorization
and exact solutions, }Phys. Lett. B, \textbf{719}, 464 (2013); K. Tuchin,
arXiv:1301.0099 [hep-ph]

\bibitem {Polikarpov}V.V. Braguta, P.V. Buividovich, T. Kalaydzhyan, M.I.
Polikarpov, \emph{Topological and magnetic properties of the QCD vacuum probed
by overlap fermions,} arXiv:1302.6458v2 [hep-lat]

\bibitem {Stern}A. Stern, Phys. Rev. Lett. \textbf{100}, 061601 (2008); T.C.
Adorno, D. M. Gitman, A.E. Shabad, D.V. Vassilevich, Phys.Rev. D \textbf{84},
085031 (2011), \textit{ibid} \textbf{84}, 065003 (2011); T.C. Adorno, D.M.
Gitman, A.E. Shabad, Phys. Rev. D \textbf{86}, 027702 (2012)

\bibitem {trudy}A. E. Shabad, \emph{Polarization of the vacuum and quantum
relativistic gas in an external field}. Nova Science Publishers, New York,
(1991). [Trudy Fiz. Inst. im. P.N. Lebedeva \textbf{192}, 5 (1988)].

\bibitem {Gies}\emph{Probing the Quantum Vacuum}, Edited by Dittrich, W.;
Gies, H., Springer Tracts in Modern Physics, vol. 166 (2000)

\bibitem {Kuznetsov}A. Kuznetsov and N. Micheev, \emph{Electroweak Processes
in External Electromagnetic Fields, }Springer Tracs in Modern Physics, vol.
197 (Springer, New York, Berlin etc., 2004)

\bibitem {GitShab}D. M. Gitman, A. E. Shabad,\emph{ Nonlinear (magnetic)
correction to the field of a static charge in an external field}. Phys. Rev. D
\textbf{86}, 125028 (2012)

\bibitem {Akhiezer}A.I. Akhiezer and V.B. Berestetskii, \emph{Quantum
Electrodynamics }(NAUKA, Moscow, 1969 (in Russian); Interscience Pbls., John
Wiley \& Sons, New York-London-Sydney, 1965)

\bibitem {BerLifPit}V.B. Berestetsky, E.M. Lifshits, and L.P. Pitayevsky,
\emph{Quantum Electrodynamics} (Nauka, Moscow, 1989; Pergamon Press Oxford,
New York, 1982)

\bibitem {Beskin}V. S. Beskin, A. V. Gurevich and Ya. N. Istomin,
\emph{Physics of the Pulsar Magnetosphere} (Cambridge, 1993)

\bibitem {hadron}T.Tatsumi, \emph{Magnetic instability of quark matter, }Phys.
Lett. B\textbf{489}, 280 (2000); \emph{Ferromagnetic properties of quark
matter -- an origin of magnetic field in compact stars --,} arXiv :0910.1642
[hep-ph] (2009)

\bibitem {weinberg}{S. Weinberg, \emph{The Quantum Theory of Fields},
(University Press, Cambridge, 2001})

\bibitem {FGS}E.S. Fradkin, D.M. Gitman and S.M. Shvartsman, \emph{Quantum
Electrodynamics with Unstable Vacuum }(Springer, Berlin, 1991)

\bibitem {Shabus}A.E. Shabad and V.V. Usov, \emph{Effective Lagrangian in
nonlinear electrodynamics and its properties of causality and unitarity,
}Phys. Rev. D \textbf{83, }105006 (2011)

\bibitem {Nakamura}J. Beringer et al. (Particle Data Group), Phys. Rev.
D\textbf{ 86}, 010001 (2012) (URL: http://pdg.lbl.gov)

\bibitem {Lee}F.X. Lee, R. Kelly, L. Zhou, W. Wilcox, \emph{Baryon magnetic
moments in the background field methods, }Phys.Lett. \textbf{B627, }71 (2005),
arXiv: 0509067 [hep-lat]

\bibitem {Butler}M.N. Butler, M.J. Savage, and R.P. Springer, Phys. Rev. D
\textbf{49, }3459 (1994), hep-ph/9308317; L. Geng, arXiv: 1301.7815 [nucl-th]

\bibitem {Magnetar}R. Turolla, S. Zane, J.A. Pons, P. Esposito, N. Rea
\emph{Is SGR 0418+5729 indeed a waning magnetar? }arXiv:1107.5488 [astro-ph.HE]

\bibitem {Kouv}C. Kouveliotou \textit{et al., }Nature (London) \textbf{393},
235 (1998); Mereghetti S. \emph{The strongest cosmic magnets: soft gamma-ray
repeaters and anomalous X-ray pulsars}, Astronomy and Astrophysics Review,
\textbf{15}, 225 (2008)

\bibitem {Usov}V.V. Usov, Nature (London) \textbf{357}, 472 (1992)

\bibitem {Arfken}G. Arfken, \emph{Mathematical methods for physicists} (Miami
University, Academic Press, Inc., Third edition, 2006)
\end{thebibliography}
\end{document}